\documentclass[12pt]{elsarticle}
 \usepackage{subfigure}
\usepackage{amsmath}
\pdfoutput=1
\usepackage{longtable,tabularx}
\usepackage{pgfplots}
\usepackage{calc}
\usepackage{textcomp}
\usepackage{float}
\usepackage{setspace}
\usepackage{natbib}
\usepackage{graphicx}
\usepackage{placeins}
\usepackage{caption}
\usepackage{array}
\usepackage{longtable}
\usepackage{multirow}

\pgfplotsset{width=7cm,compat=1.7}
\makeatletter
\def\fixedlabel#1#2{%
  \@bsphack%
  \protected@write\@auxout{}%
         {\string\newlabel{#1}{{#2}{\thepage}}}%
  \@esphack}
\makeatother

\journal{Journal of Composite Structures}

\begin{document}

\begin{frontmatter}

\title{Analyzing Thermal Buckling in Curvilinearly Stiffened Composite Plates with Arbitrary Shaped Cutouts Using Isogeometric Level Set Method}

\author[First]{Balakrishnan Devarajan\corref{cor1}}
\ead{dbalak9@vt.edu}

\cortext[cor1]{Corresponding Author}

\address[First]{Department of Biomedical Engineering and Mechanics, Virginia Polytechnic Institute and State University, Blacksburg, VA 24061, USA}

\begin{abstract}
In this paper we develop a new simple and effective isogeometric analysis
for modeling thermal buckling of stiffened laminated composite plates with
cutouts using level sets. We employ a first order shear deformation theory to
approximate the displacement field of the stiffeners and the plate. Numerical
modeling with a treatment of trimmed objects, such as internal cutouts
in terms of NURBS-based isogeometric analysis presents several challenges,
primarily due to need for using the tensor product of the NURBS basis functions.
Due to this feature, the refinement operations can only be performed
globally on the domain and not locally around the cutout. The new approach
can overcome the drawbacks in modeling complex geometries with multiple-patches
as the level sets are used to describe the internal cutouts; while the
numerical integration is used only inside the physical domain. Results of
parametric studies are presented which show the influence of ply orientation,
size and orientation of the cutout and the position and profile of the curvilinear
stiffeners. The numerical examples show high reliability and efficiency of
the present method compared with other published solutions and ABAQUS.
\end{abstract}


\end{frontmatter}

\section{Introduction}

 %In recent times, unitized structural design has been considered to build a significantly
%lighter and more environment-friendly aircraft. The key idea of such a design is to have the stiffening members  integral to the structure, leading to a monolithic construction of aircraft components \cite{Renton2004}. 
%Advanced multi-layered composite structures are being increasingly used in aerospace, shipbuilding, bridge and other industries due to their enhanced strength to weight ratios. However, when used in  space vehicles flying at hypersonic speeds such structures
%experience significant temperature rise in very short time resulting due to aerodynamic heating
%due to friction between the vehicle surface and the atmosphere. Such phenomena is more prominent during reentry and launch processes. 

Due to a large requirement of laminates for a variety of engineering applications,  the use of plate structures with arbitrary cutouts are inevitable. The presence of cutouts can significantly affect the  behaviors of structures. Studies on the analysis of eigenvalues and stability of  laminates with cutouts are of great importance for many practical applications including airplane wing, fuselage and ribs.

In recent years, the isogeometric analysis (IGA) \cite{Hughes2005}, \cite{Yu2015} has become popular since it inherently owns many great advantages including exact geometry representation, higher-order continuity, simple mesh refinement, and avoiding the traditional mesh generation procedure. The traditional mesh generation in real world design problems could be complicated \cite{de2018structural, de2019structural,de2021lightweight,de2019unconventional, de2019structural,jrad2017global,de2021algorithms} Many insights into mathematical properties \cite{Bazilevs2008}, \cite{Evans2009a}, integration method \cite{Hughes2010a}, \cite{Auricchio2012a} and splines techniques \cite{Bazilevs2010} have been gained. The inherent characteristics of IGA make it superior to the classical FEM in many aspects \cite{Hughes2005}. The IGA has successfully applied to many engineering problems including plates and shells \cite{Valizadeh2013},  \cite{yin2014isogeometric}, \cite{kapoor2013interlaminar}, \cite{kapoor2012geometrically} fluid mechanics \cite{Bazilevs2008},  damage and fracture mechanics \cite{Verhoosel2011a}, contact mechanics \cite{Lu2011}, unsaturated flow problem in porous media \cite{nguyen2014isogeometric}, and structural shape optimization \cite{Wall2008}.

The higher-order smoothness of the NURBS basis functions have huge benefits with regards to the analysis of plates and shells. Based on the Kirchhoff plate theory, construction of rotation-free isogeometric shell was first proposed \cite{kiendl2009isogeometric}, and then developed for multiple NURBS patches by Kiendl \cite{Kiendl2010}. Later, this element was extended for vibration and buckling analyses of laminated plates \cite{shojaee2012free},  functionally graded plates \cite{yin2014isogeometric}, bending and buckling analysis of laminated composite plates \cite{kapoor2012geometrically,kapoor2013interlaminar,tran2014isogeometric} and cloth simulation \cite{Lu2011}.\par  However, most of the problems investigated were simple single-patch geometry. In modeling a more complex geometry like a plate with a clover shaped internal cutout a trimmed NURBS surface can be useful. A NURBS surface and a set of ordered boundary curves lying within the parameter space of the surface \cite{Piegl1998} can be used to describe a trimmed NURBS surface. Shojaee et al. \cite{shojaee2012free,Shojaee2012a} divided a rectangular plate with a heart shaped cutout into multiple NURBS patches, and to ensure continuity between adjacent patches, the authors applied the bending strip method. However this approach is quite cumbersome to implement and is also observed to complicate the unification of design and analysis, which is the sole aim of IGA. Trimmed  NURBS surfaces were handled by Schmidt et al. \cite{Schmidt2012} using a local reconstruction technique, which can handle trimmed NURBS surfaces under an isogeometric analysis framework. The method involved reconstructing trimmed NURBS patches using a least square approximation or by an interpolation method having known the trimming curves. The method is effective and can be easily applied to an arbitrarily shaped area on the surface. However, the method could not exactly describe trimmed boundaries and the number of sampling points heavily influenced the accuracy of the reconstructed boundaries. 
In order to model complex geometry problems, the IGA has been used in combination with an enrichment function similar to XFEM, to develop XIGA \cite{Ghorashi2012a}. It has also been used in combination with the finite cell method (FCM) \cite{Rank2012} and \cite{Schillinger2012a}. In this method, internal cutouts were described using level sets. However, level set methods using IGA were developed mostly for elastostatic problems, and are yet to be implemented for thermal buckling analysis of stiffened laminated plates with complicated internal cutouts. The objective of this research, is to fill this gap. 

Thermal effects, especially thermal buckling is important to be considered in design, analysis and weight optimization of aircraft structures\cite{devarajan2016thermomechanical}and \cite{2016-01-1986}. Such problems have been researched by many in the past. For instance, Noor and Burton \cite{noor1996computational},\cite{Burton1994} and \cite{noor1992three} used 3D elasticity to analyze the thermal buckling of multilayered and sandwich plates. However, it is difficult to solve problems which involve complicated geometry under different boundary conditions. The problem becomes even more challenging if the plate is reinforced with stiffeners. \par
Several authors in the past have studied the behavior of such stiffened plates. A finite element algorithm was developed by Mukhopadhyay and Mukherjee \cite{mukhopadhyay1990finite} to study the buckling of stiffened panels under uniaxial compression. Prusty and Satsangi \cite{prusty2001finite} used the same algorithm to perform structural analysis of stiffened isotropic plates and shells. While the aforementioned authors studied straight stiffeners, curvilinearly stiffened panels were  analyzed by Tamijani and Kapania \cite{Tamijani2010}, Zhao and Kapania \cite{Zhao2016}, De and Kapania \cite{de2019structural,robinson2016aeroelastic,de2017sparibs,de2017structural} and Shi et al. \cite{Shi2015}. It has been found by Hao et al. \cite{hao2016optimization} that curvilinear stiffeners if used can improve the strength margins of panels since they reinforce the cutouts in them. The tension field resulting due to the presence of curvilinear stiffeners was examined and the loading path was compared with results obtained using straight stiffeners.

In this work, thermal buckling analysis of stiffened laminated composite plates with a cutout is presented. Level sets are used to describe the cutouts and numerical integration is used only inside the physical domain. The plate is assumed to be subjected to a uniform temperature rise for different boundary conditions. The stiffness matrices and thermal force vector are derived according to first-order shear deformation theory (FSDT). Firstly, the present formulation is verified by comparison with available references from literature. The significant effects of aspect ratio, length-to-thickness ratio, fiber orientation and boundary condition on the behavior of composite plates are then analyzed and useful conclusions are derived from the numerical results.

\section{Mathematical foundations}
\subsection{Laminated composite panel}

\begin{figure}[htbp]
  \centering
 \includegraphics[width=1\textwidth]{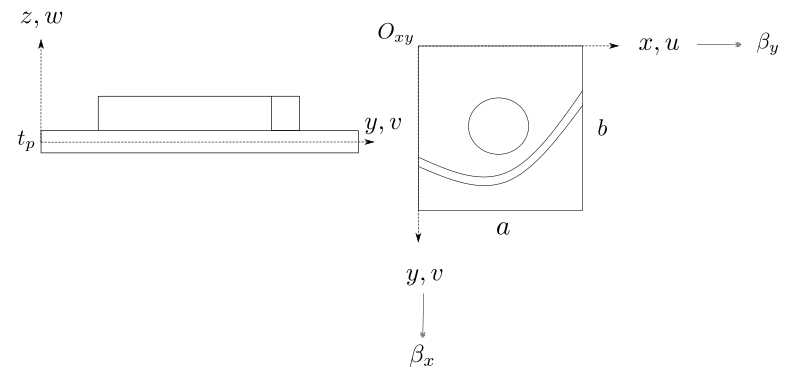}
  \caption{Geometry and nomenclature of a stiffened composite panel with a central cutout.}
\label{fig:ls_plate_with_a_cutout}
\end{figure}

According to the first-order shear deformation theory (FSDT) the displacement field is
considered as the first-order Taylor series expansion of the mid-plane displacement variables with respect to plate thickness as follows:
Consider a stiffened composite plate with a circular cutout as shown in Figure ~\ref{fig:ls_plate_with_a_cutout}.
The mid-plane of the panel $O_{xy}$ is used as the reference plane for the global coordinate system. The plate is assumed to have a uniform thickness with no plydrops. Let $u$ be the displacements in the $x$-axis and $v$ be the in-plane displacements along the $y$-axis. Let $w$ be the transverse deflection along the $z$-axis.

\FloatBarrier  \FloatBarrier

Let $\beta_{y}$ and $\beta_{x}$ be rotation components of the panel around the $x$- and $y$-axes, respectively. The displacement components of the panel can now be written as,
\begin{equation}
\begin{array}{cc}&u\left(x\text{,}y\text{,}z\text{,}t\right)=u_0\left(x\text{,}y\text{,}z\text{,}t\right)+z\beta_x\left(x\text{,}y\text{,}t\right)\\&v\left(x\text{,}y\text{,}z\text{,}t\right)=v_0\left(x\text{,}y\text{,}z\text{,}t\right)+z\beta_y\left(x\text{,}y\text{,}t\right)\\&w\left(x\text{,}y\text{,}z\text{,}t\right)=w_0\left(x\text{,}y\text{,}z\text{,}t\right)\end{array}
\end{equation}

The plate strain energy $U_p$ can be written as,
\begin{equation}
    U_p=\frac12\underset\Omega{\hspace{0.28em}\int\int\hspace{0.28em}}\boldsymbol \varepsilon_\mathbf p^{\mathbf L^{\mathrm T}}{\mathbf D}_\mathbf p\boldsymbol \varepsilon_\mathbf p^\mathbf L\mathrm d\Omega
\end{equation}

Where, $\mathbf D_{p}$ depends on the material property, stacking sequence, ply orientation and thickness of the plate. Derivation of $\mathbf D_{p}$  is explained in Section \ref{ss:mp_Orthotropic Layers}.
The generalized strains $\boldsymbol \varepsilon_\mathbf p^\mathbf L$  can be expressed in terms of the generalized displacements $\mathbf {u_{p}}$  of the panel as,
\begin{equation}
\boldsymbol \varepsilon_\mathbf p^\mathbf L=
\begin{Bmatrix} 
\varepsilon_{x}^{0} \\ 
\varepsilon_{y}^{0}  \\ 
\gamma_{xy}^{0}  \\ 
\kappa_{x}^{0} \\
\kappa_{y}^{0} \\
\kappa_{xy}^{0} \\
\gamma_{xz}^{0} \\
\gamma_{yz}^{0} \\
\end{Bmatrix}=\begin{bmatrix} 
\frac{\partial}{\partial{x}} & 0  & 0 & 0 & 0 \\ 
0 & \frac{\partial}{\partial{x}}  & 0 & 0 & 0\\ 
\frac{\partial}{\partial{x}} & \frac{\partial}{\partial{x}}  & 0 & 0 & 0\\ 
0 & 0  & 0 & \frac{\partial}{\partial{x}} & 0\\ 
0 & 0  & 0 & 0 & \frac{\partial}{\partial{x}}\\ 
0 & 0  & 0 & \frac{\partial}{\partial{x}} & \frac{\partial}{\partial{x}}\\ 
0 & 0  & \frac{\partial}{\partial{x}} & 1 & 0 \\
0 & 0  & \frac{\partial}{\partial{x}} & 0 & 1 \\
\end{bmatrix}\begin{Bmatrix} 
u_{0} \\ 
v_{0}  \\ 
w_{0}  \\ 
\beta_{x} \\
\beta_{y}
\end{Bmatrix}=\mathbf {B_{p}} \mathbf {u_{p}} 
\end{equation}
\newline
The strain energy of the composite panel $U_{p}$ is,
\begin{equation}
U_{p}=\frac{1}{2}\iint_{\Omega}^{}
\mathbf{u_{p}}
^T
\mathbf{B_{p}}
^T
\mathbf{D_{p}}
\mathbf{B_{p}}
\mathbf{u_{p}}
d\Omega
\end{equation}
For thermal bucking analysis, the  geometric stiffness due to inplane stress resultants is considered. Thus, the potential
$W_p$ of the panel due to in-plane stress vector ${\overline\sigma}_\mathbf p={\left(\sigma_x^0\hspace{0.28em}\hspace{0.28em}\hspace{0.28em}\sigma_y^0\hspace{0.28em}\hspace{0.28em}\hspace{0.28em}\tau_{xy}^0\hspace{0.28em}\hspace{0.28em}\hspace{0.28em}0\hspace{0.28em}\hspace{0.28em}\hspace{0.28em}0\right)}$ is,

\begin{equation}
\label{eq:birds3}
W_p=\frac12\underset\Omega{\hspace{0.28em}\int\int\hspace{0.28em}}{\overline\sigma}_\mathbf p\boldsymbol \varepsilon_\mathbf p^\mathbf{NL}\mathrm d\Omega
\end{equation}

The nonlinear strain factor matrix is:

\begin{equation}
\left[\mathbf B_p^\mathit{NL}\right]=\begin{bmatrix}0&0&\frac\partial{\partial x}&0&0\\0&0&\frac\partial{\partial y}&0&0\\0&0&0&\frac\partial{\partial x}&0\\0&0&0&\frac\partial{\partial y}&0\\0&0&0&0&\frac\partial{\partial x}\\0&0&0&0&\frac\partial{\partial y}\end{bmatrix}
\end{equation}

The in-plane stress matrix can be written as:
\begin{equation}
{\boldsymbol\sigma}_\mathbf p={\left(\begin{array}{cccccc}t\sigma_y^0&t\tau_{xy}^0&0&0&0&0\\t\tau_{xy}^0&t\sigma_y^0&0&0&0&0\\0&0&\frac{t^3}{12}\sigma_y^0&\frac{t^3}{12}\tau_{xy}^0&0&0\\0&0&\frac{t^3}{12}\tau_{xy}^0&\hspace{0.28em}\frac{t^3}{12}\sigma_y^0&0&0\\0&0&0&0&\frac{t^3}{12}\sigma_y^0&\frac{t^3}{12}\tau_{xy}^0\\0&0&0&0&\frac{t^3}{12}\tau_{xy}^0&\hspace{0.28em}\frac{t^3}{12}\sigma_y^0\end{array}\right)},
\end{equation}
Hence, Equation \ref{eq:birds3} can be written as 
\begin{equation}
   W_p =\frac12\underset\Omega{\hspace{0.28em}\int\int\hspace{0.28em}}\mathbf u_\mathbf p^\mathbf T\mathbf B_\mathbf p^{\mathbf{NL}^{\mathrm T}}{\mathbf\sigma}_\mathbf p\mathbf B_\mathbf p^\mathbf{NL}{\mathbf u}_\mathbf p\mathrm d\Omega
\end{equation}
\subsection{Orthotropic layers}
\label{ss:mp_Orthotropic Layers}
The constitutive equation in the local coordinate system for the $k^{th}$ orthotropic elastic lamina can be derived from Hooke\textquotesingle s law as
\begin{equation}
\begin{Bmatrix}\sigma_1\\\sigma_2\\\tau_{12}\\\sigma_3\\\tau_{31}\\\tau_{23}\end{Bmatrix}^{(k)}=\begin{bmatrix}Q_{11}&Q_{12}&0&0&0&0\\Q_{21}&Q_{22}&0&0&0&0\\0&0&Q_{66}&0&0&0\\Q_{31}&Q_{32}&0&Q_{33}&0&0\\0&0&0&0&Q_{55}&0\\0&0&0&0&0&Q_{44}\end{bmatrix}^{(k)}\begin{Bmatrix}\varepsilon_1-\alpha_1\mathrm\Delta T\\\varepsilon_2-\alpha_2\mathrm\Delta T\\\gamma_{12}\\\varepsilon_3-\alpha_3\mathrm\Delta T\\\gamma_{31}\\\gamma_{23}\end{Bmatrix}^{(k)}
\end{equation}
where $Q_{ij}$  are the elastic coefficients in the material coordinate system,  $T$ is the temperature change and $\alpha_i$  is the thermal coefficient of expansion in the principal $i^{th}$-direction.
The  transformed material constants are expressed as,
\begin{equation}
\begin{Bmatrix}{\overline Q}_{11}\\{\overline Q}_{12}\\{\overline Q}_{22}\\{\overline Q}_{16}\\{\overline Q}_{26}\\{\overline Q}_{66}\end{Bmatrix}=\begin{bmatrix}c^4&2c^2s^2&s^4&4c^2s^2\\c^2s^2&c^4+s^4&c^2s^2&-4c^2s^2\\s^4&2c^2s^2&c^4&4c^2s^2\\c^3s&\mathit{cs}(s^2-c^2)&-\mathit{cs}^3&2\mathit{cs}(s^2-c^2)\\\mathit{cs}^3&\mathit{cs}(c^2-s^2)&-c^3s&2\mathit{cs}(c^2-s^2)\\c^2s^2&-2c^2s^2&c^2s^2&{(c^2-s^2)}^2\end{bmatrix}\begin{Bmatrix}Q_{11}\\Q_{12}\\Q_{22}\\Q_{66}\end{Bmatrix}\text{,}
\end{equation}
\begin{equation}
\begin{bmatrix}{\overline Q}_{44}&{\overline Q}_{13}\\{\overline Q}_{45}&-{\overline Q}_{36}\\{\overline Q}_{55}&{\overline Q}_{23}\end{bmatrix}=\begin{bmatrix}c^2&s^2\\-\mathit{cs}&\mathit{cs}\\s^2&c^2\end{bmatrix}\begin{bmatrix}Q_{44}&Q_{13}\\Q_{55}&Q_{23}\end{bmatrix}\text{,}\hspace{0.35em}{\overline Q}_{33}=Q_{33}
\end{equation}
\begin{equation}
\begin{Bmatrix}\alpha_x\\\alpha_y\\\alpha_\mathit{xy}\end{Bmatrix}=\begin{bmatrix}c^2&s^2\\s^2&c^2\\2\mathit{cs}&-2\mathit{cs}\end{bmatrix}\begin{Bmatrix}\alpha_1\\\alpha_2\end{Bmatrix}\text{,}\hspace{0.35em}\alpha_z=\alpha_3
\end{equation}
where $c= \cos\theta$ and $s= \sin\theta$   .
% \subsection{Weak form equation}
% A weak form of the buckling analysis for laminated composite plates can be briefly expressed as:
% \begin{equation}
% \int_\Omega\delta\mathbf\varepsilon_p^T\mathbf D^\ast{\mathbf\varepsilon}_p\hspace{0.12em}\mathrm d\Omega+\int_\Omega\delta\mathbf\gamma^T\mathbf D_S^\ast\mathbf\gamma\hspace{0.12em}\mathrm d\Omega+h\int_\Omega\nabla^T\delta w{\oversetˆ{\mathbf\sigma}}_0\nabla w\hspace{0.12em}\mathrm d\Omega=0\text{,}
% \end{equation}
% where $\rho$ is the transverse loading per unit area and strain components ${\mathbf\varepsilon}_p$ and $\gamma$ are expressed by
% \begin{equation}
% {\mathbf\varepsilon}_p=\{{\mathbf\varepsilon}_0\hspace{1em}{\mathbf\kappa}_1\hspace{1em}{\mathbf\kappa}_2\}^T\text{,}\hspace{1em}\boldsymbol\gamma=\{{\mathbf\varepsilon}_s\hspace{1em}{\mathbf\kappa}_s\}^T
% \end{equation}
\newline Using this, $\mathbf{D}_p$ can now be written as 

\begin{equation}
\mathbf {D}_p=\begin{bmatrix}\mathbf A&\mathbf B&\mathbf 0\\\mathbf B&\mathbf D&\mathbf 0\\\mathbf 0&\mathbf 0&\mathbf A^\text{s}\end{bmatrix}
\end{equation}
where,
\begin{equation}
(A_\mathit{ij}\text{,}{B}_\mathit{ij}\text{,}D_\mathit{ij})=\int_{-h/2}^{h/2}\overline Q_\mathit{ij}(1\text{,}z\text{,}z^2)\text{d}z\text{,}\hspace{1em}A_\mathit{ij}^\text{s}=K\int_{-h/2}^{h/2}\overline Q_\mathit{ij}\hspace{0.12em}\text{d}z
\end{equation}
where $A_\mathit{ij}$, $B_\mathit{ij}$ and $D_\mathit{ij}$ are valid for $i, j = 1, 2, 6$, and $A_\mathit{ij}^\text{s}$ for $i, j = 4$, 5 according to the Voigt notation. $K$ denotes the transverse shear correction coefficient. A value of $K=5/6$ was used for the analyses.

\subsection{Curvilinear stiffener}
\label{ss:stiffmodel}
\begin{figure}[htbp]
\centering
   \includegraphics[width=1\textwidth]{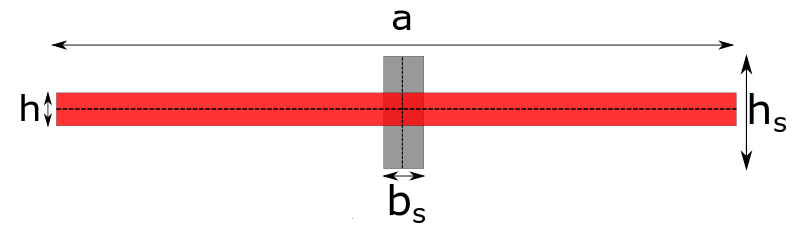}
  \caption{Composite plate (red) stiffened by a rectangular stiffener (gray)}
         \label{fig:mp_stiffener}
\end{figure}

Consider a stiffener attached to a
plate as shown in Figure \ref{fig:mp_stiffener}.
\FloatBarrier  \FloatBarrier
The width and height of the rectangular stiffener are given as $b_s$ and $h_s$, respectively. Details of the coordinate system and nomenclature can be found in \cite{Zhao2016}. The stiffener is modeled
using 1D $2^{nd}$ order NURBS basis function. The stiffener is modeled as a Timoshenko beam. A detailed description of how displacement compatibility between the control points of the stiffener and the plate was achieved can be found in \cite{devarajan2020thermal} and \cite{devarajan2019thermomechanical}.

\subsection{Isogeometric analysis}
\label{ss:mp_isogeom}
The concept of IGA is to use the same basis functions to generate CAD and to discretize the solution to the boundary value problem. This isoparametric formulation in combination with the properties of NURBS basis functions allows the geometry to be modeled exactly from the coarsest level of refinement, and maintains this exact representation as the mesh is refined. The elements are constructed by a knot span which is an array of non-decreasing sequence of parameter values \cite{Hughes2005}. The NURBS basis functions are used to construct the exact geometry, as well as the corresponding solution space. In this section, we present a primer on NURBS and also some
basic formulations of IGA.
Knot vectors can be defined as:
\begin{equation}
\Xi =\{\xi_1, \xi_2, ..., \xi_{n+p+1}\}
\end{equation}
B-spline basis functions are defined recursively starting with piecewise constants $p=0$
\begin{equation}
N_i^0(\xi)=\left\{\begin{array}{ccc}1&\mathit{if}&\xi_i<\xi<\xi_{i+1}\\0&&\text{otherwise}\end{array}\right.
\end{equation}
For higher values like $p=0,1,2...$ the basis functions are defined using a recursion relation
\begin{equation}
N_i^p(\xi)=\frac{\xi-\xi_i}{\xi_{i+p}-\xi_i}N_i^{p-1}(\xi)+\frac{\xi_{i+p+1}-\xi}{\xi_{i+p+1}-\xi_{i+1}}N_{i+1}^{p-1}(\xi)
\end{equation}
where $\xi_1$ is the $i^{th}$ knot, $n$ is the number of basis functions and $p$ is the polynomial order. \\
We denote Non-uniform Rational B-splines (NURBS) functions  expressed as
\begin{equation}
R_{i,p}(\xi)=\frac{N_{i,p}(\xi)w_i}{\sum_{j=1}^nN_{j,p}(\xi)w_j}
\end{equation}
Where $w_{i}$ is the weight corresponding to the $i^{th}$ control point.
For a 2-dimensional surface, the basis functions are expressed as:
\begin{equation}
R_{i\text{,}j}^{p\text{,}q}(\xi\text{,}\eta)=\frac{N_{i\text{,}p}(\xi)N_{j\text{,}q}(\eta)w_{i\text{,}j}}{\sum_{i=1}^n\sum_{j=1}^mN_{i\text{,}p}(\xi)N_{j\text{,}q}(\eta)w_{i\text{,}j}}=\frac{N_{i\text{,}p}(\xi)N_{j\text{,}q}(\eta)w_{i\text{,}j}}{W(\xi\text{,}\eta)}
\end{equation}
The 2D surface is generated using the basis functions, $R_{i,j}$, and the control points 
\textbf{$P_{i,j}$}
\begin{equation}
\boldsymbol S(\xi\text{,}\eta)=\sum_{i=1}^n\sum_{j=1}^mR_{i\text{,}j}^{p\text{,}q}(\xi\text{,}\eta){\boldsymbol P}_{i\text{,}j}\text{,}
\end{equation}
Further details can be found in \cite{Hughes2005} and \cite{tran2014isogeometric}.

 \subsection{Level Set Function}
A signed distance function is utilized here to identify the enriched elements (elements cut by the cutout boundary), the outer elements (elements that are part of the plate) and inner elements (elements that fall completely within the cutout geometry) using the equation of a circle, given as:
  \begin{equation}
    \phi(\mathbf{X})=  |\mathbf{X}-\mathbf{X_{c}}|-r
\end{equation}

Where $\phi(\mathbf{X})$ is the level set value at coordinates $\mathbf{X}$. $\mathbf{X_{c}}$ is the coordinates of the center of the
circular cutout and $r$ is the radius of the cutout. These values are stored at all control points
for later use. Next, the coordinates of the points where the geometry of the cutout intersects the element is computed (Figure \ref{fig:triangulationtwoimages}). This process will approximate the shape of the cutout as a polygon with sides equal to the number of enriched elements. This process has no impact on elements that are not cut by the interface.

The implemented enrichment function is:
  \begin{equation}
    \psi(\mathbf{X})= \sum_{I=1}^{4} |\phi_{I}|N(\mathbf{X})_{I}-|\sum_{I=1}^{4} \phi_{I}N(\mathbf{X})_{I}|
\end{equation}

The enrichment function $\psi$ at coordinates $\mathbf{X}$ is evaluated as a combination of control point level
set values and the shape functions. Compared to other choices for enrichment functions
this one has two major advantages. As the enrichment function has a zero value outside the cut
elements as well as in the cut element control points there will be no contribution to the global stiffness
matrix from an element that are not cut by the interface and the displacements evaluated at
the control points will not have any extra contribution either.

\subsection{Triangulation and Numerical Integration}
\label{ss:Appendix}
\begin{figure}[htbp]
\centering
  \includegraphics[width=1.1\textwidth]{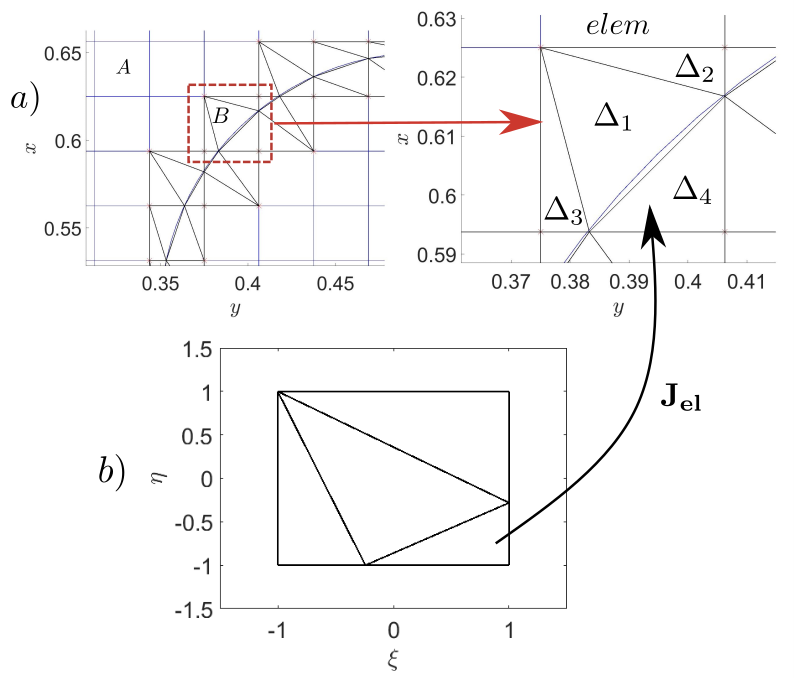}
\caption{ Mappings and transformation from physical space to natural space}
\label{fig:triangulationtwoimages}
\end{figure}

Figure \ref{fig:triangulationtwoimages} shows a non enriched element (A) and an enriched element (B) in the $x-y$ coordinate system. Since some part of such enriched elements lie within the cutout, for numerical integration purposes, they have to be treated differently. For this, the coordinates of the points where the geometry of the cutout intersects the element is computed (Figure \ref{fig:triangulationtwoimages}). The creation of additional points will enable the division of such a quadrilateral element into triangles and the process will approximate the shape of the cutout as a polygon with sides equal to the number of enriched elements

Consider one such triangular geometry ($\Delta_{1}$). Let the  function $f(x,y)$ determine some
property at $[x, y]$, a point within this geometry. Suppose another property, that is evaluated by the integration
of $f(x,y)$ over this geometry is of interest. An analytical calculation of the integrand is either
impossible or very expensive. It is convenient to use Gauss quadrature formula that approximate the integral as a weighted sum of the function values at a number
of predetermined points. Note that the Gauss integration computed using $n$ points is exact for a function of order $2n-1$ in 1D. These points are given in a transformed coordinate system (natural coordinate system). \\ Thus, to integrate $f(x,y)$ over $\Delta_{1}$,
\begin{equation}
    I=\int_{\Delta_{1}}^{} f(x,y)dxdy
\end{equation}
a transformation into this natural coordinate system and a change of variables is necessary. This transformation is performed using a Jacobian matrix,
$\mathbf{J_{el}}$ and is defined by

\begin{equation}
\begin{bmatrix}
dx\\ 
dy
\end{bmatrix}=\mathbf{J_{el}} \begin{bmatrix}
d\xi  \\ 
d\eta 
\end{bmatrix}
\end{equation}
Where,
\begin{equation}
    \mathbf{J_{el}} = \begin{bmatrix}
\frac{\partial x}{\partial \xi } & \frac{\partial x}{\partial \eta  }\\ 
\frac{\partial y}{\partial \xi } & \frac{\partial y}{\partial \eta  }
\end{bmatrix}
\end{equation}

\begin{figure}[htbp]
\centering
  \includegraphics[width=0.6\textwidth]{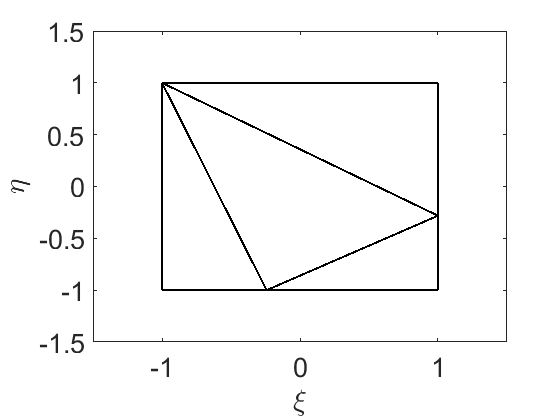}
\caption{Element and triangle transformed into coordinate system appropriate for $B_{matrix}$
calculation
}
\label{fig:triang2}
\end{figure}

Then it follows that
\begin{equation}
    I =\frac{1}{2}\int_{0}^{1}\int_{0}^{1-\eta} f(x(\xi,\eta),y(\xi,\eta))\begin{vmatrix}
\frac{\partial x}{\partial \xi } & \frac{\partial x}{\partial \eta  }\\ 
\frac{\partial y}{\partial \xi } & \frac{\partial y}{\partial \eta  }
\end{vmatrix} d\xi d\eta
\end{equation}
By numerical integration the quantity  can be evaluated
\begin{equation}
    I \cong \frac{1}{2} \sum_{1}^{i} w_{i} f(x(\xi,\eta),y(\xi,\eta))\begin{vmatrix}
\frac{\partial x}{\partial \xi } & \frac{\partial x}{\partial \eta  }\\ 
\frac{\partial y}{\partial \xi } & \frac{\partial y}{\partial \eta  }
\end{vmatrix}
\end{equation}
Where $i$ is the number of Gauss points in the triangle.
This will be useful in  codes where element stiffness matrices for elements with
enriched nodes are evaluated.
\FloatBarrier  \FloatBarrier
The element stiffness matrix will have a shape as follows:

\begin{equation}
  \mathbf{K_{e}^{ij}} =\iint_{elem} \mathbf{B_{i}^{T}}\mathbf{D}\mathbf{B_{j}}tdxdy  
\end{equation}
Where, 
\begin{equation}
\mathbf{B_{i}}=[
\mathbf{B_{STD}} \ | \
\mathbf{B_{ENR}}]
\end{equation}
Where $\mathbf{B_{i}}$ could be one of the three matrices, $\mathbf{B_{p}}$, $\mathbf{B_{s}}$ or $\mathbf{B_{p}^{NL}}$. The subscripts $\mathbf{STD}$ and $\mathbf{ENR}$ refer to standard non-enriched elements and enriched elements  respectively. For a four noded isoparametric element the shape functions are defined in the isoparametric
domain. However, the $\mathbf{B}$ matrices are functions of the shape functions which are defined in the isoparametric coordinate system $\xi-\eta$ (Figure \ref{fig:triang2}) , but the Gauss points and weights associated for a triangular domain aren't. 

\FloatBarrier  \FloatBarrier 

\begin{figure}[htbp]
\centering
  \subfigure[3 Gauss points]{\includegraphics[width=3.1in]{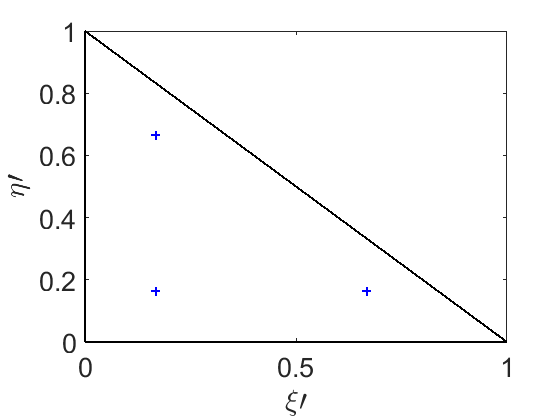}}
  \subfigure[7 Gauss points]{\includegraphics[width=3.1in]{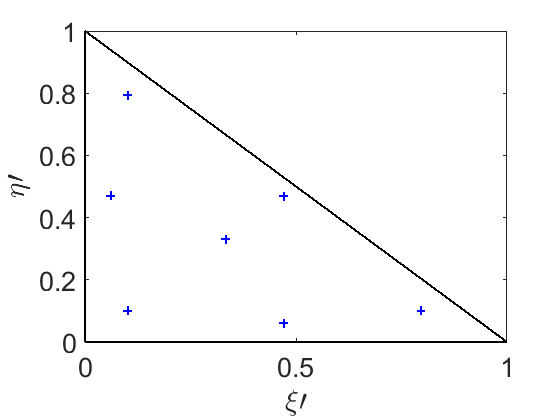}}
    		\caption{Triangle transformed into coordinate system compatible with given Gauss-points}\label{fig:triang3}
\end{figure}

They are defined in the $\xi'-\eta'$ coordinate system (Figure \ref{fig:triang3} with the vertices of the triangle located at:
\begin{equation}
\left [  \xi{}_{1}',\eta_{1}'\right ]=\left [0 \ 0\right ];\left [  \xi{}_{2}',\eta_{2}'\right ]=\left [1 \ 0\right ];\left [  \xi{}_{1}',\eta_{1}'\right ]=\left [0 \ 1 \right ];
\end{equation}
Hence, the mapping of these Gauss points to the isoparametric coordinate system is to be performed for each triangle.

 These Gauss points can be mapped using nodal shape functions and isoparametric coordinates of vertices as follows:

\begin{equation}
\begin{bmatrix}
\xi{}^{}_{gp} & \eta{}^{}_{gp}
\end{bmatrix}_{\Delta_{1}}=
    \begin{bmatrix}
1-\xi{}^{'}_{gp}-\eta{}^{'}_{gp}\\ 
\xi{}^{'}_{gp}\\ 
\eta{}^{'}_{gp}
\end{bmatrix}^{T}_{\Delta_{1}} \begin{bmatrix}
\xi_{1} & \eta_{1}\\ 
\xi_{2} & \eta_{2}\\ 
\xi_{3} & \eta_{3}\\ 
\end{bmatrix}_{\Delta_{1}}
\end{equation}

The weight associated with each Gauss point will be scaled using the ratio of the areas of the triangles in the $\xi$-$\eta$ coordinate system and the $\xi'$-$\eta'$ coordinate system. We now have Gauss points and associated weights in the isoparametric coordinate system to in order to perform numerical integration.

\FloatBarrier  \FloatBarrier

Now, all the tools needed for the numerical integration are available.
The stiffness matrix contribution from the triangle can be calculated now as
\begin{equation}
 \iint_{\Delta } \mathbf{B_{i}^{T}}(x,y)\mathbf{D}\mathbf{B_{j}}(x,y)tdxdy  =  \frac{1}{2} \sum_{k=1}^{n}  \mathbf{B_{i}}(x,y)^{T}\mathbf{D}\mathbf{B}_{j}(x,y)tw_{k}\left |\mathbf{J}_{\Delta }  \right | \left |\mathbf{J}_{el}  \right |
\end{equation}

Where $B_{i}$ and $B_{j}$ are either $\mathbf{B_{STD}}$ or $\mathbf{B_{ENR}}$, depending on what part of the stiffness matrix
are currently being calculated and $n$ is the number of Gauss points in the element.

The target expression (a sample entry in a $\mathbf{B_{STD}}$ matrix) is,
\begin{equation}
    N_{1,x}=\frac{\partial \xi}{\partial x}\frac{\partial N_{1}}{\partial \xi}+\frac{\partial \eta}{\partial x}\frac{\partial N_{1}}{\partial \eta}
\end{equation}

Where $N$ is the basis function.
While, for the enriched B-matrix $\mathbf{B_{ENR}}$,
\begin{equation}
    \frac{\partial (\psi N_{1} )}{\partial x}= N_{1}\frac{\partial \psi  }{\partial x}+\frac{\partial  N_{1} }{\partial x} \psi
\end{equation}

Where, $\psi$ is the enrichment function. 
 \subsection{A Three point parametrization of the curvilinear stiffener} \label{ss:ls_multipatchsec}
Previous researchers \cite{Tamijani2010}, \cite{Zhao2016} and \cite{Shi2015} modeled curvilinear stiffeners using a parabolic curve. Using the parametric equation of the parabola and the start and the end point coordinates, one can obtain the three control points which could define the Bezier curve \cite{Piegl1996}. Since any parabolic stiffener can be represented using three control points, the coordinates of the end control points $\Delta\epsilon$ and Point 2 (where  $\delta_{dist}$  refers to the point [$\delta_{dist}$ ,$\delta_{dist}$]) were used to represent the curvilinear stiffener.
\FloatBarrier \begin{figure}[htbp]
\centering
  \includegraphics[width=0.5\textwidth]{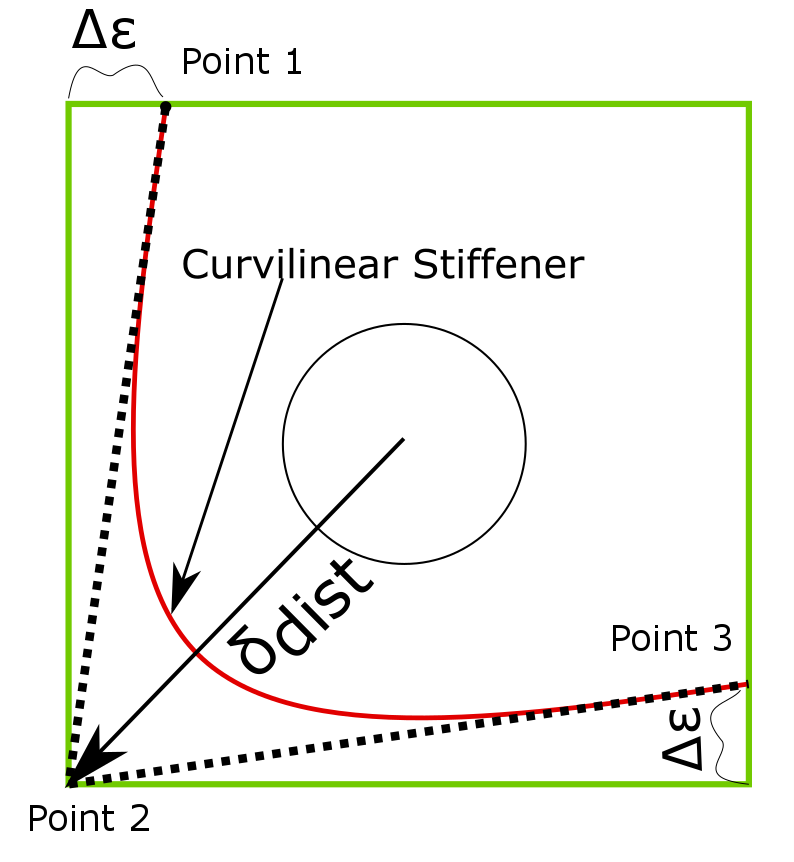}
\caption{Parametrization of curvilinear stiffener}
\label{fig:ls_Parametrization}
\end{figure} \FloatBarrier
\section{Results and Discussion} \label{se:ls_one_section}
To demonstrate and  validate the method, thermal buckling analysis of curvilinearly stiffened laminated composite plates with various cutouts is carried out. Results are compared with existing literature for some cases while for some others, using commercially popular software like ABAQUS. Parametric studies were performed which show the influence of both the cutout profile and the shape and position of the curvilinear stiffeners in the results.

\subsection{Thermal buckling of laminated composite plates}
Over the years, and most recently \cite{Tran2017}, several researchers have analyzed the thermal buckling of symmetric laminated plates. A four-layer ${[}0^{\circ}/90^{\circ}/90^{\circ}/0^{\circ}{]}$ laminated plate with the two ratios of $a/h = 10$ (Figure \ref{fig:mp_stiffener}) and $100$ were investigated for \cite{babu2000refined} two sets of boundary conditions. The plate is assumed to be a square of edge length 1 unit to make normalizations easy.

\begin{equation}
CCCC: \hspace{0.35em}
u_0=0,v_0=0,\psi_x=0,\psi_y=0, w_0=0
\end{equation}
on all four edges.\newline 
\begin{equation}
\mathit{SSSS}:\left\{\begin{array}{c}u_0=v_0=w_0=\phi_y=\phi_z=0\hspace{1em}\text{on}\hspace{0.35em}x\hspace{0.35em}=\hspace{0.35em}0\text{,}\hspace{0.35em}a\\u_0=v_0=w_0=\phi_x=\phi_z=0\hspace{1em}\text{on}\hspace{0.35em}y\hspace{0.35em}=\hspace{0.35em}0\text{,}\hspace{0.35em}b\end{array}\right.
\end{equation}
Material properties are defined as \cite{Kant2000}, \cite{babu2000refined} and\cite{Tran2017}
\begin{equation}
\begin{array}{c}E_L/E_T=15\text{,}\hspace{0.35em}G_\mathit{LT}/E_T=0.5\text{,}\hspace{0.35em}G_\mathit{TT}/E_T=0.3356\text{,}\\\nu_\mathit{LT}=0.3\text{,}\hspace{0.35em}\nu_\mathit{TT}=0.49\text{,}\hspace{0.35em}\alpha_L/\alpha_0=0.015\text{,}\hspace{0.35em}\alpha_T/\alpha_0=1\end{array}\end{equation}
where $L$ and $T$ refer to the directions parallel and perpendicular to the fibers and $\alpha_0$ is the normalization factor for the thermal expansion coefficient. Table
\ref{tab:ls_compositenostiff} illustrates the comparison of critical temperature obtained by NURBS-based IGA and FEM-Q16 \cite{Kant2000}, \cite{babu2000refined} and\cite{Tran2017}.
The critical buckling temperature values of the thinner plate have been scaled by 100.

\subsection{Isotropic plate with a circular hole at the center}
In order to illustrate the performance of IGA code while modeling plates with holes, an isotropic plate with a circular hole at the center was considered. Variation of the critical thermal load with respect to the radius of the hole was compared against Avci et al.\cite{Avci2005}. The results were obtained for simply supported (SSSS) and clamped (CCCC) boundary conditions. The material properties of the plate are given as:
\begin{equation}
\begin{array}{c}E\hspace{0.35em} =\hspace{0.35em} 208\hspace{0.35em} GPa\text{,}\hspace{0.35em}\nu =\hspace{0.35em} 0.3\text{,}\hspace{0.35em}\alpha =\hspace{0.35em} 1.17e-5 \hspace{0.35em} ^\circ C^{-1} \end{array}\end{equation}
As seen from the Table \ref{tab:ls_2}, the results are in good agreement with \cite{Avci2005}.
\FloatBarrier \begin{table}[htbp]										\caption	{Variation	of	critical	load	with	respect	to	radius 	of	the	circular	cutout}	\label{tab:ls_2}
\centering															
\begin{tabular}{|c|c|c|c|c|}															
															
\hline															
	&	\multicolumn{2}{c|}{CCCC}	&	\multicolumn{2}{c|}{SSSS}	\\	\hline									
$\frac{diameter}{width}$	&	Present	&	Avci	&	Present	&	Avci	\\	\hline					
0	&	29.35	&	29.16	&	10.83	&	10.97	\\	\hline					
0.1	&	28.01	&	28.57	&	10.64	&	10.82	\\	\hline					
0.2	&	29.17	&	29.07	&	10.52	&	10.68	\\	\hline					
0.3	&	35.27	&	35.37	&	11.01	&	11.14	\\	\hline					
0.4	&	49.53	&	49.67	&	12.48	&	12.57	\\	\hline					
0.5	&	78.74	&	78.55	&	15.41	&	15.34	\\	\hline					
\end{tabular}															
\end{table} \FloatBarrier

\subsection{Composite plate with elliptical cutout}
\label{ss:ls_4.3}
\begin{figure}[htbp]
\centering  
\includegraphics[width=.5\textwidth]{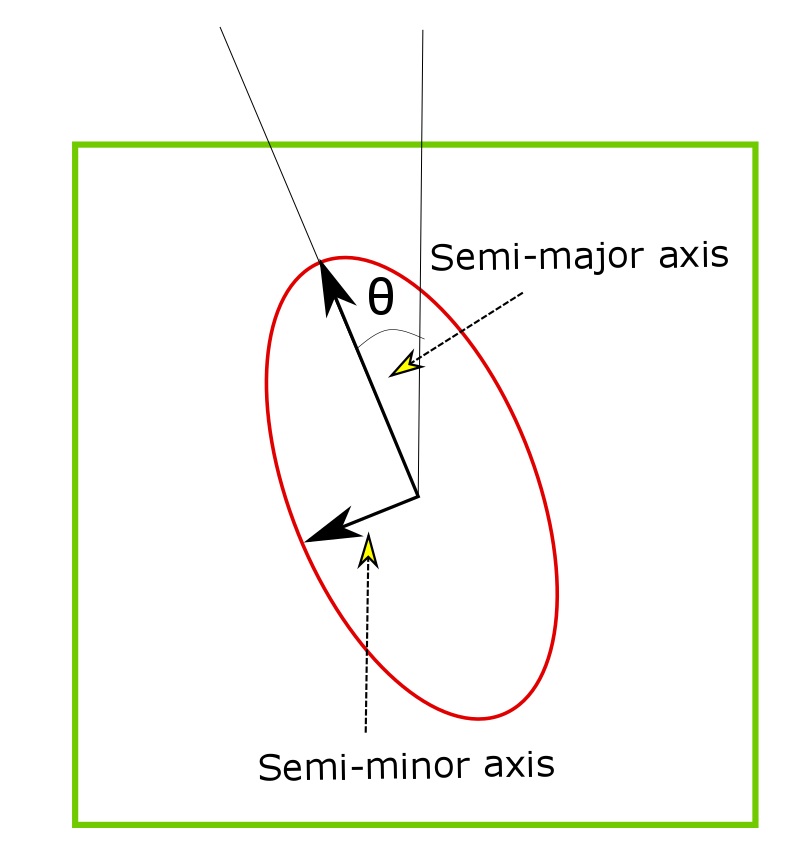}
\caption{Dimensions and orientation of an elliptical cutout}
\label{fig:ls_Parametrize}
\end{figure}

To demonstrate the versatility and to validate the method, buckling analysis of plates with an elliptical cutout was considered (Figure \ref{fig:ls_Parametrize}). Results are compared with those obtained using ABAQUS. A unit square domain is considered with an elliptical cutout. The selected examples are for comparison purpose only and do not represent
the full capabilities of the proposed method.
\FloatBarrier  \FloatBarrier
The ellipse is defined by the semi major axis, the semi minor axis and the angle of orientation $\theta$. For validation purposes, an ellipse of semi-major axis = 0.2, semi-minor axis = 0.1 and $\theta = 0$ was considered.

\begin{figure}[htbp]
\centering
  \subfigure[ABAQUS]{\includegraphics[width=2in]{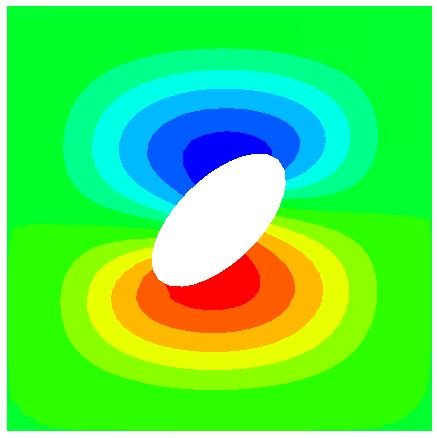}}
  \subfigure[IGA]{\includegraphics[width=2in]{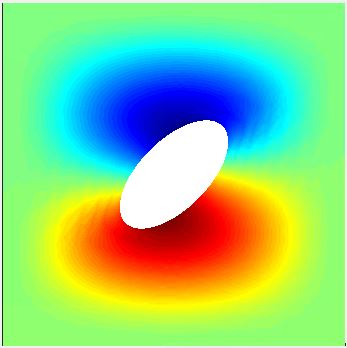}}

    		\caption{Buckling mode shape plots of a plate with an elliptical cutout ($\theta = 45^{\circ} $)}\label{fig:ls_modeplot}
\end{figure}

\FloatBarrier \begin{table}[htbp]
\centering
\caption{The convergence of the critical thermal buckling load of a four-layer $[0^{\circ}/90^{\circ}/90^{\circ}/0^{\circ}]$ laminated composite square plate.}
\begin{tabular}{|c|c|c|}
\hline
\textbf{Number of elements} & \textbf{ABAQUS} & \textbf{Present} \\ \hline
\textbf{16}                 & 0.636           & 0.653        \\ \hline
\textbf{64}                 & 0.406           & 0.421       \\ \hline
\textbf{256}                & 0.3782          & 0.401        \\ \hline
\textbf{1024}               & 0.377           & 0.381        \\ \hline
\end{tabular}
\label{tab:ls_convergence}
\end{table} \FloatBarrier
It can be observed from Table \ref{tab:ls_convergence} that level set IGA approach yields relatively the same rate of convergence as ABAQUS. For mode shape comparison (Figure \ref{fig:ls_modeplot}) purposes, a plate with an elliptical cutout ($\theta = 45^{\circ} $) was modeled in ABAQUS with an
irregular mesh composed of 1028 elements. S8R:  An 8-node doubly curved thick shell, reduced integration element was used to mesh the domain. IGA mode shape plot was obtained using 1024 elements. 

\FloatBarrier  \FloatBarrier

%\FloatBarrier \begin{figure}[htbp]	
	%\centering
	%\begin{subfigure}[t]{3in}
	%	\centering
	%	\includegraphics[width=3in]{modeplotabaqus.jpg}
	%	\caption{ABAQUS}		
	%\end{subfigure}
	%\quad
	%\begin{subfigure}[t]{3in}
	%	\centering
	%	\includegraphics[width=3in]{modeplot.jpg}
	%	\caption{IGA}
	%\end{subfigure}
    %		\caption{Buckling mode shape plots of a plate with an elliptical cutout ($\theta = 45^{\circ} $)}\label{fig:ls_modeplot}
%\end{figure} \FloatBarrier
Next, the variation of critical buckling load with respect to change in eccentricity and angle of orientation of the cutout was analyzed. Table \ref{tab:ls_ellipiso} presents the results for three different eccentricities. The results of the present method (LSM) are compared with those obtained using the Multi-Patch method (MPM)

\FloatBarrier \begin{table}[htbp]
\centering
\caption{Variation of thermal buckling load of a four-layer $[0^{\circ}/90^{\circ}/90^{\circ}/0^{\circ}]$ laminated composite square plate.}
\begin{tabular}{|c|c|c|c|c|c|c|}
\hline
\textbf{$\theta$} & \multicolumn{2}{c|}{\textbf{\begin{tabular}[c]{@{}c@{}}Semi-minor axis=0.1 \\ Semi-major axis=0.2\end{tabular}}} & \multicolumn{2}{c|}{\textbf{\begin{tabular}[c]{@{}c@{}}Semi-minor axis=0.1 \\ Semi-major axis=0.3\end{tabular}}} & \multicolumn{2}{c|}{\textbf{\begin{tabular}[c]{@{}c@{}}Semi-minor axis=0.2 \\ Semi-major axis=0.3\end{tabular}}} \\ \hline
 & \textbf{LSM} & \textbf{MPM} & \textbf{LSM} & \textbf{MPM} & \textbf{LSM} & \textbf{MPM} \\ \hline
\textbf{0} & 0.381  & 0.379 & 0.413 & 0.402 & 0.633 & 0.633 \\ \hline
\textbf{15} & 0.395  & 0.390 & 0.443  & 0.438 & 0.659 & 0.653 \\ \hline
\textbf{30} & 0.427 & 0.421 & 0.532 & 0.525 & 0.729 & 0.715 \\ \hline
\textbf{45} & 0.469  & 0.460 & 0.641 & 0.630 & 0.809  & 0.807 \\ \hline
\end{tabular}
\label{tab:ls_ellipiso}
\end{table} \FloatBarrier

\subsection{Curvilinearly stiffened composite panels with a noncentric circular cutout}
\label{ss:ls_noncencutout}
\begin{figure}[htbp]

\begin{longtable}{ >{\centering\arraybackslash} m{0.25cm} >{\centering\arraybackslash} m{4cm} >{\centering\arraybackslash} m{0.7cm} >{\centering\arraybackslash} m{4cm} >{\centering\arraybackslash} m{0.7cm}}
\hline 
 & \multicolumn{2}{c}{Present}    & \multicolumn{2}{c}{ABAQUS}    \\
\hline \\
Mode & & $\Delta T_{cr}$ & & $\Delta T_{cr}$\\ \hline
 1  & \includegraphics[scale=.31]{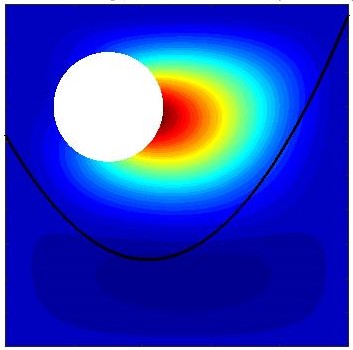} & 0.426 &  \includegraphics[scale=.16]{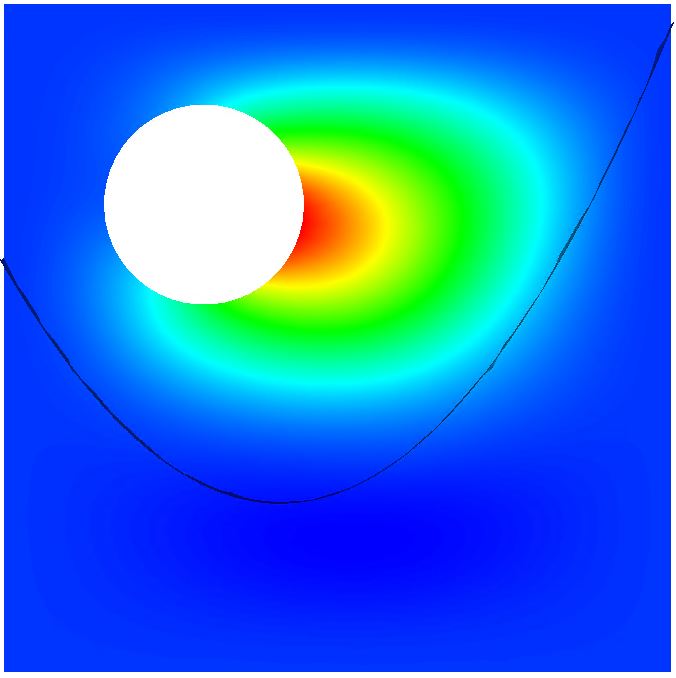} & 0.420\\
 2 & \includegraphics[scale=.31]{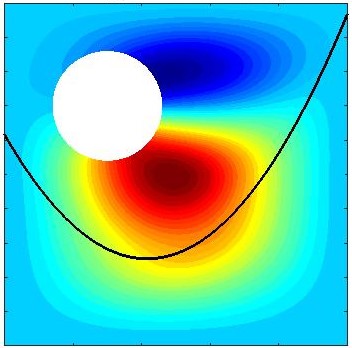} & 0.535 &  \includegraphics[scale=.16]{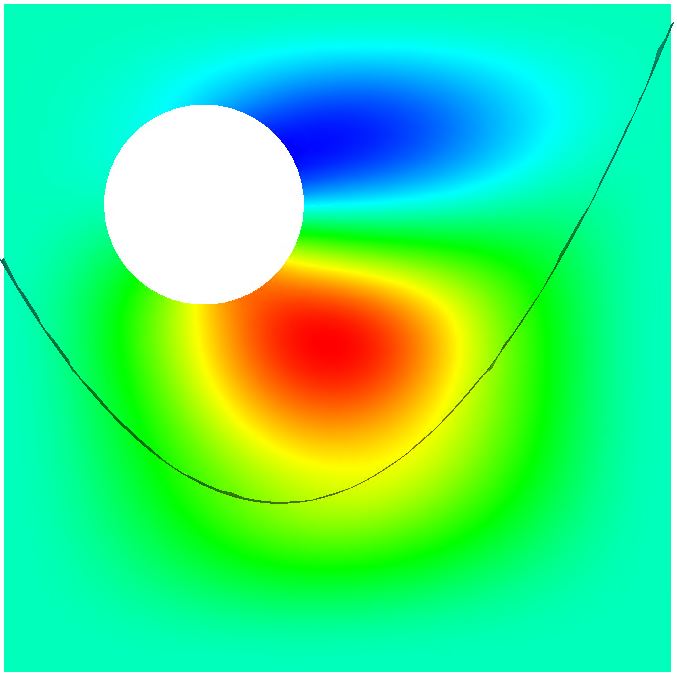} & 0.523\\
 3 & \includegraphics[scale=.31]{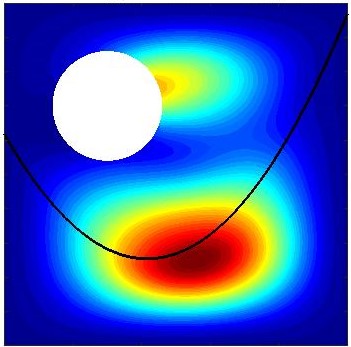} & 0.708 &  \includegraphics[scale=.16]{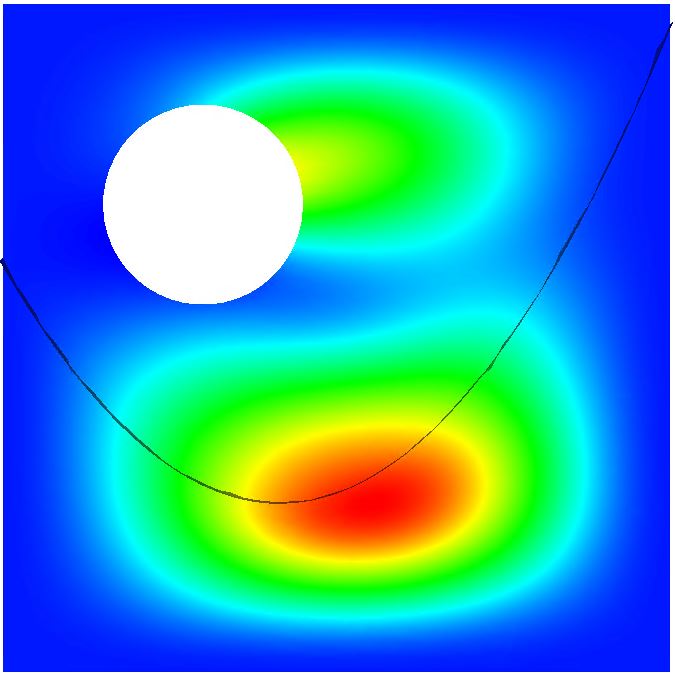} & 0.698\\
 4 & \includegraphics[scale=.31]{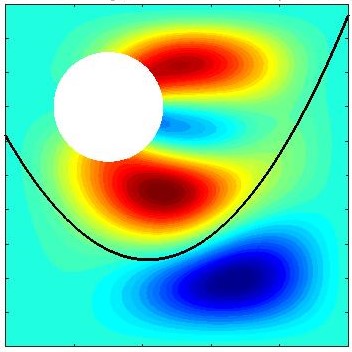} & 0.910 &  \includegraphics[scale=.16]{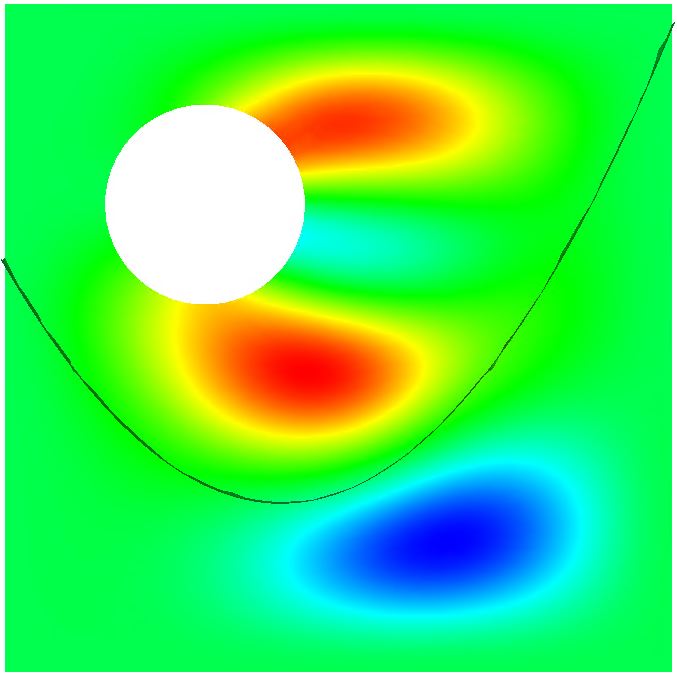} & 0.906\\
 5 & \includegraphics[scale=.31]{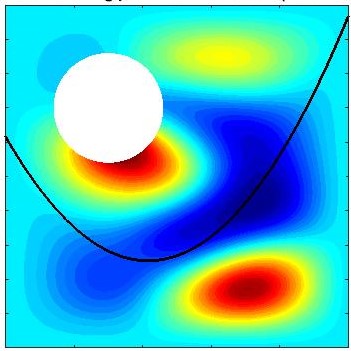} & 1.070 &  \includegraphics[scale=.16]{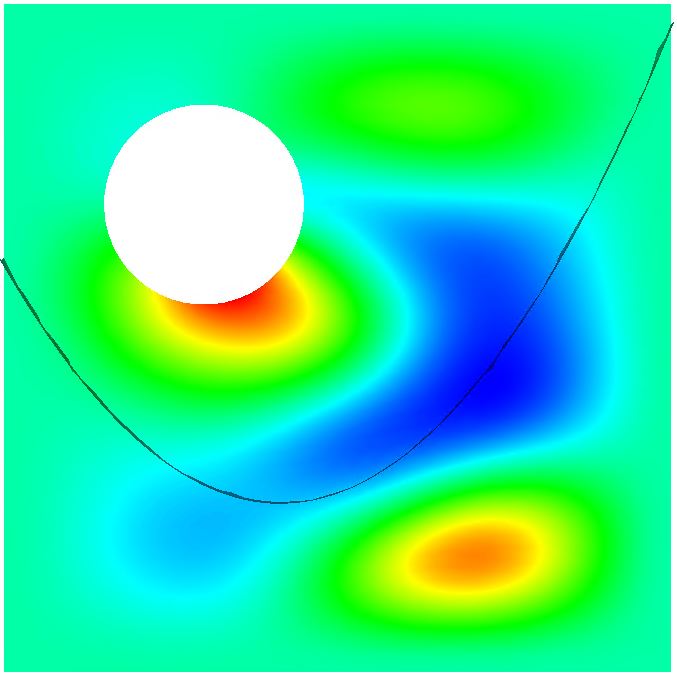} & 1.007\\
\endfirsthead
\endhead

\end{longtable}
\caption{First five eigenmode shape plots for a plate with curvilinear stiffener using  LSM and ABAQUS, a commercial
available software.}
\label{fig:ls_modeplotcomp}
\end{figure}

In this sub section, the influence of the noncentric circular cutouts on clamped stiffened composite panels is studied. The stiffener is isotropic  with Young's modulus = $E_{T}$, Poisson's ratio = $\nu_{LT}$ and coefficient of thermal expansion = $\alpha_{0}$. The stiffness ratio $\gamma=\mathit{EI}/\mathit{bD}$  and the area ratio $\delta=A_s/{\mathit{bt}}_p$ are 5 and 0.1 respectively unless specified. To check the robustness of the level set method, results for clamped curvilinearly stiffened composite plate are compared with those obtained
using ABAQUS. A noncentric circular cutout of radius $0.15$ with center at ($0.3, 0.7$) was assumed. The curvilinear stiffener configuration was adopted from \cite{Tamijani2010} and scaled proportionally to match the dimensions of the plate. From the ABAQUS library, the plate was meshed using 3040 S8R elements and curvilinear stiffener was meshed using 192 elements S8R elements (48 divisions along the arc length and 4 divisions along the depth direction). The IGA plate model was meshed using 4096 elements and the curvilinear stiffener using 32 elements. The critical buckling temperature result was observed to converge for this mesh configuration. The first five eigen buckling models were extracted for each case. As seen from Figure \ref{fig:ls_modeplotcomp} the buckling mode shapes obtained using IGA and ABAQUS are comparable.

\pagebreak

\FloatBarrier  \FloatBarrier

\subsection{Curvilinearly stiffened composite panels with an elliptical central cutout}

In this section, results of curvilinearly stiffened composite panels with elliptical cutouts are presented. An ellipse of semi-major axis = 0.2 and semi-minor axis = 0.1  is considered for all cases. The  variation of the critical thermal load with respect to angle orientations, ply orientations, stiffness ratios and different stiffener profiles (defined by $\Delta\epsilon$ Figure \ref{fig:ls_Parametrization}) are presented in Table \ref{tab:ls_9} and Table \ref{tab:ls_8}. The results obtained using present method (LSM) and the Multi-Patch method (MPM) are in excellent agreement. 

%\FloatBarrier \begin{table}[htbp]
%\caption {Variation of critical buckling load with respect to angle of orientation of the elliptical cutout for different ply orientations for $\gamma=5$} \label{tab:ls_9} 
%\centering
%\begin{tabular}{|c|c|c|c|c|}
%\hline
%\multicolumn{5}{|c|}{\textbf{$\Delta\epsilon$ = 0}}                                                                       \\ \hline
%\textbf{$\theta$} & \textbf{${[}0^{\circ}/90^{\circ}/0^{\circ}/90^{\circ}{]}$} & \textbf{${[}0^{\circ}/90^{\circ}/90^{\circ}/0^{\circ}{]}$} & \textbf{${[}45^{\circ}/-45^{\circ}/-45^{\circ}/45^{\circ}{]}$} & \textbf{${[}45^{\circ}/-45^{\circ}/45^{\circ}/-45^{\circ}{]}$} \\ \hline
%\textbf{0}  & 0.479 & 0.646 & 0.414 & 0.666 \\ \hline
%\textbf{15} & 0.502 & 0.685 & 0.393 & 0.671 \\ \hline
%\textbf{30} & 0.548 & 0.721 & 0.371 & 0.655 \\ \hline
%\textbf{45} & 0.599 & 0.737 & 0.362 & 0.644 \\ \hline
%\end{tabular}
%\end{table} \FloatBarrier

\FloatBarrier \begin{table}[htbp]
\centering
\caption {Variation of critical buckling load with respect to angle of orientation of the elliptical cutout for different ply orientations for $\gamma=5$} \label{tab:ls_9}
\begin{tabular}{|c|c|c|c|c|c|c|c|c|}
\hline
\multicolumn{9}{|c|}{\textbf{ $\Delta\epsilon=0$}}                                                                       \\ \hline
\textbf{$\theta$} & \multicolumn{2}{c|}{\textbf{\begin{tabular}[c]{@{}c@{}}Antisymmetric\\ Cross-Ply\end{tabular}}} & \multicolumn{2}{c|}{\textbf{\begin{tabular}[c]{@{}c@{}}Symmetric\\ Cross-Ply\end{tabular}}} & \multicolumn{2}{c|}{\textbf{\begin{tabular}[c]{@{}c@{}}Symmetric\\ Angle-Ply\end{tabular}}} & \multicolumn{2}{c|}{\textbf{\begin{tabular}[c]{@{}c@{}}Antisymmetric\\ Angle-Ply\end{tabular}}} \\ \hline
 & \textbf{LSM} & \textbf{MPM} & \textbf{LSM} & \textbf{MPM} & \textbf{LSM} & \textbf{MPM} & \textbf{LSM} & \textbf{MPM} \\ \hline
\textbf{0}	&	0.475	&	0.473	&	0.413	&	0.406	&	0.425	&	0.423	&	0.456	&	0.455	\\	\hline
\textbf{45}	&	0.507	&	0.497	&	0.494	&	0.488	&	0.370	&	0.366	&	0.436	&	0.434	\\	\hline
\multicolumn{9}{|c|}{\textbf{ $\Delta\epsilon=0.25$}} \\ \hline
\textbf{0}	&	0.559	&	0.553	&	0.492	&	0.481	&	0.469	&	0.462	&	0.548	&	0.542	\\	\hline
\textbf{45}	&	0.598	&	0.592	&	0.595	&	0.590	&	0.406	&	0.395	&	0.525	&	0.521	\\	\hline
\end{tabular}
\end{table} \FloatBarrier

%\FloatBarrier \begin{table}[htbp]
%\caption {Variation of critical buckling load with respect to the angle of orientation of the elliptical cutout for different ply orientations for $\gamma=5$} \label{tab:ls_8} 
%\centering
%\begin{tabular}{|c|c|c|c|c|}
%\hline
%\multicolumn{5}{|c|}{\textbf{ $\Delta\epsilon$ = 0.25}}                                                                       \\ \hline
%\textbf{$\theta$} & \textbf{${[}0^{\circ}/90^{\circ}/0^{\circ}/90^{\circ}{]}$} & \textbf{${[}0^{\circ}/90^{\circ}/90^{\circ}/0^{\circ}{]}$} & \textbf{${[}45^{\circ}/-45^{\circ}/-45^{\circ}/45^{\circ}{]}$} & \textbf{${[}45^{\circ}/-45^{\circ}/45^{\circ}/-45^{\circ}{]}$} \\ \hline
%\textbf{0}                     & 0.685              & 0.998                    & 0.750                        & 0.969                        \\ \hline
%\textbf{15}                    & 0.729              & 1.104                    & 0.720                        & 0.990                        \\ \hline
%\textbf{30}                    & 0.827              & 1.180                    & 0.695                        & 1.002                        \\ \hline
%\textbf{45}                    & 0.911              & 1.244                    & 0.682                        & 0.989                        \\ \hline
%\end{tabular}
%\end{table} \FloatBarrier
\FloatBarrier \begin{table}[htbp]
\centering
\caption {Variation of critical buckling load with respect to the angle of orientation of the elliptical cutout for different ply orientations for $\gamma=10$} \label{tab:ls_8} 
\begin{tabular}{|c|c|c|c|c|c|c|c|c|}
\hline
\multicolumn{9}{|c|}{\textbf{ $\Delta\epsilon=0$}}                                                                       \\ \hline
\textbf{$\theta$} & \multicolumn{2}{c|}{\textbf{\begin{tabular}[c]{@{}c@{}}Antisymmetric\\ Cross-Ply\end{tabular}}} & \multicolumn{2}{c|}{\textbf{\begin{tabular}[c]{@{}c@{}}Symmetric\\ Cross-Ply\end{tabular}}} & \multicolumn{2}{c|}{\textbf{\begin{tabular}[c]{@{}c@{}}Symmetric\\ Angle-Ply\end{tabular}}} & \multicolumn{2}{c|}{\textbf{\begin{tabular}[c]{@{}c@{}}Antisymmetric\\ Angle-Ply\end{tabular}}} \\ \hline
 & \textbf{LSM} & \textbf{MPM} & \textbf{LSM} & \textbf{MPM} & \textbf{LSM} & \textbf{MPM} & \textbf{LSM} & \textbf{MPM} \\ \hline
\textbf{0}	&	0.488	&	0.485	&	0.417	&	0.409	&	0.443	&	0.443	&	0.469	&	0.468	\\	\hline
\textbf{45}	&	0.509	&	0.509	&	0.502	&	0.495	&	0.83	&	0.378	&	0.448	&	0.449	\\	\hline
\multicolumn{9}{|c|}{\textbf{ $\Delta\epsilon=0.25$}} \\ \hline
\textbf{0}	&	0.590	&	0.581	&	0.509	&	0.501	&	0.485	&	0.479	&	0.576	&	0.572	\\	\hline
\textbf{45}	&	0.619	&	0.612	&	0.618	&	0.610	&	0.410	&	0.405	&	0.560	&	0.554	\\	\hline
\end{tabular}
\end{table} \FloatBarrier

\subsection{Curvilinearly stiffened composite panels with a complicated cutout}
\begin{figure}[htbp]
\centering
  \includegraphics[scale=0.2]{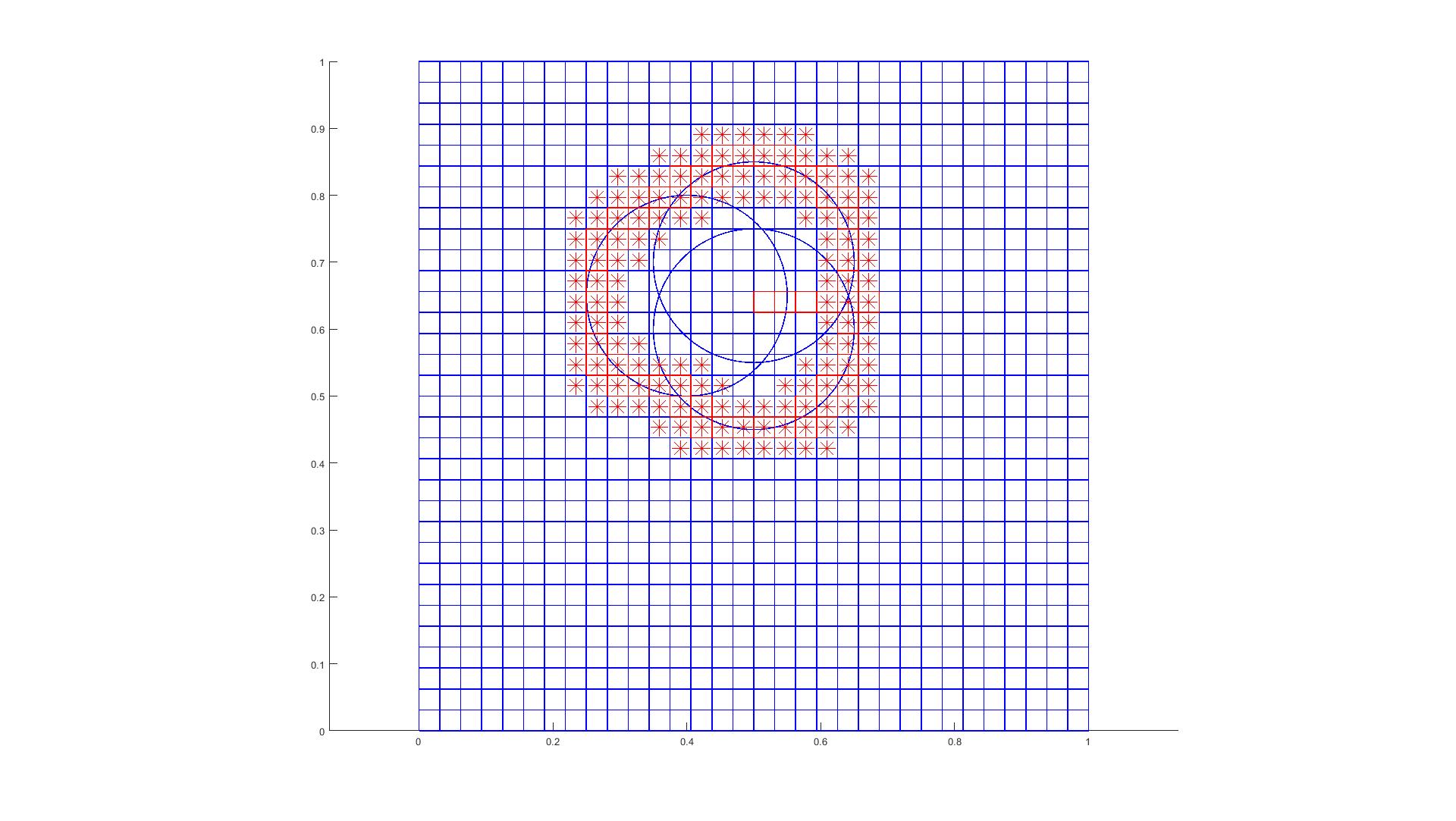}
\caption{Isogeometric mesh plot of a plate with clover shaped cutout constructed with three cirles. The enriched elements are shown in red}
\label{fig:ls_complicatedmesh}
\end{figure}

The present method is applied to analyze a stiffened laminated composite plate with a clover shaped cutout Figure \ref{fig:ls_complicatedmesh}. The cutout is a boolean of three circles of radius $0.15$ resulting in a clover shape. The center of circles \textbf{A}, \textbf{B} and \textbf{C} are  at ($0.4, 0.65$), ($0.5, 0.7$) and ($0.5, 0.6$) respectively. Figure \ref{fig:ls_complicatedmesh} shows the mesh plot for the plate geometry with the red region being the enriched elements.\par
\begin{figure}[htbp]

\begin{longtable}{ >{\centering\arraybackslash} m{0.25cm} >{\centering\arraybackslash} m{4cm} >{\centering\arraybackslash} m{0.7cm} >{\centering\arraybackslash} m{4cm} >{\centering\arraybackslash} m{0.7cm}}
\hline 
 & \multicolumn{2}{c}{Present}    & \multicolumn{2}{c}{ABAQUS}    \\
\hline \\
Mode & & $\Delta T_{cr}$ & & $\Delta T_{cr}$\\ \hline
 1 & \includegraphics[scale=.31]{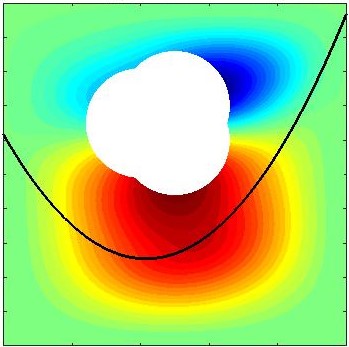} & 0.751 & \includegraphics[scale=.20]{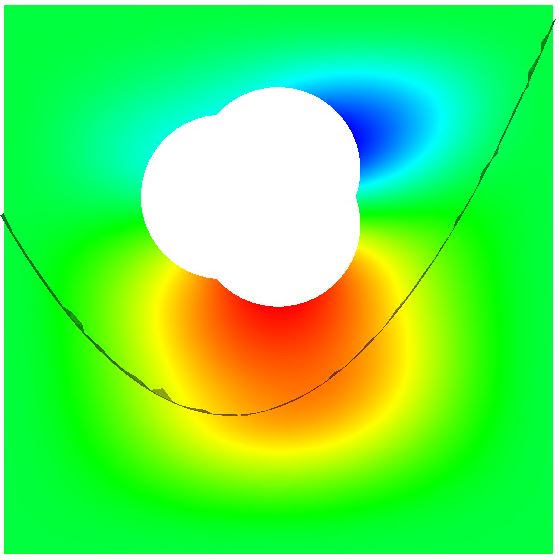} & 0.748\\
 2 & \includegraphics[scale=.31]{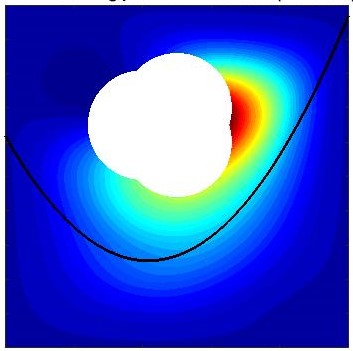} & 0.827 & \includegraphics[scale=.20]{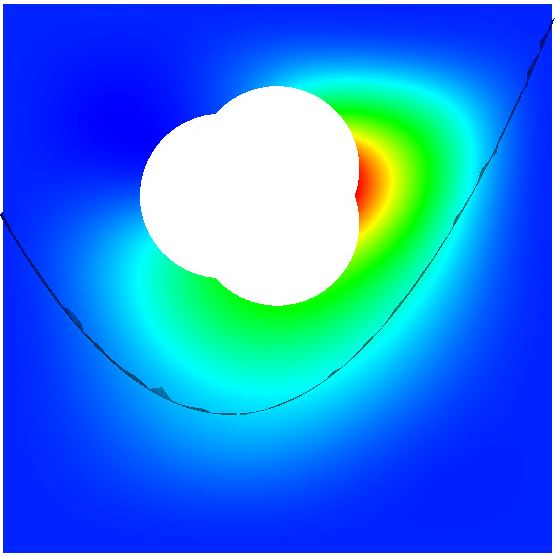} & 0.817 \\
 3 & \includegraphics[scale=.31]{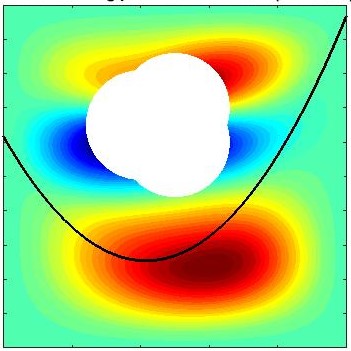} & 0.961 & \includegraphics[scale=.20]{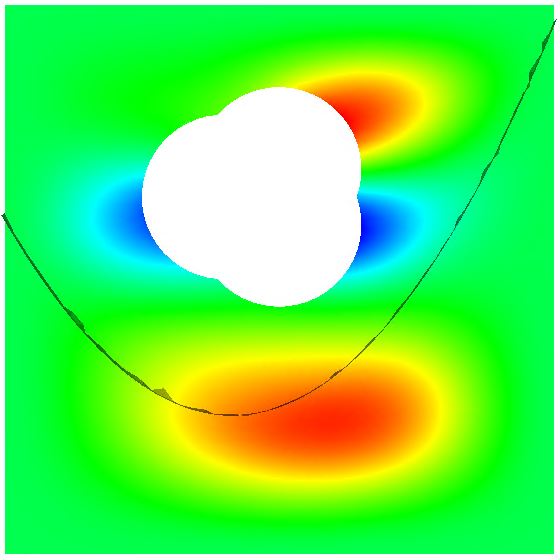} & 0.955\\
 4 & \includegraphics[scale=.31]{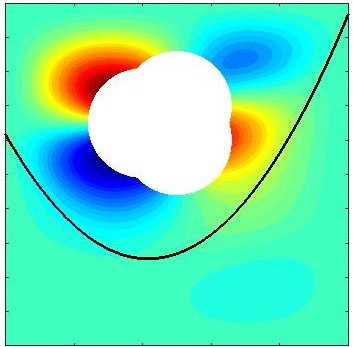} & 1.009 & \includegraphics[scale=.20]{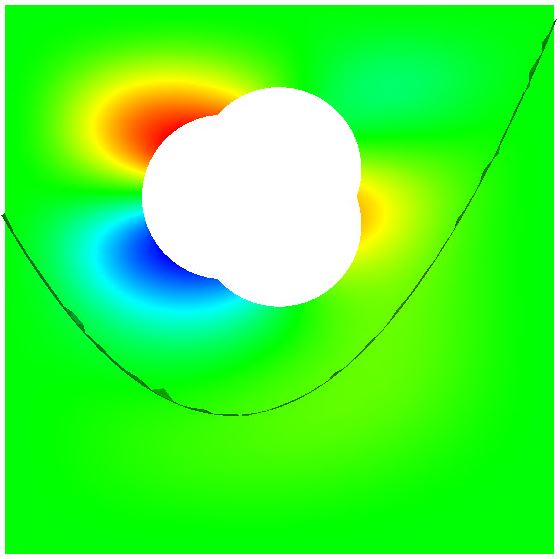} & 1\\
 5 & \includegraphics[scale=.31]{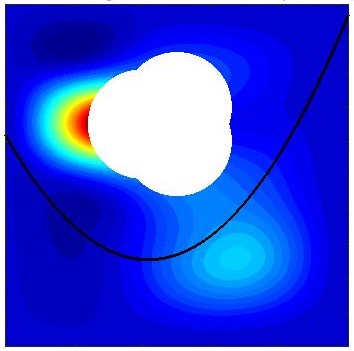} & 1.096 & \includegraphics[scale=.20]{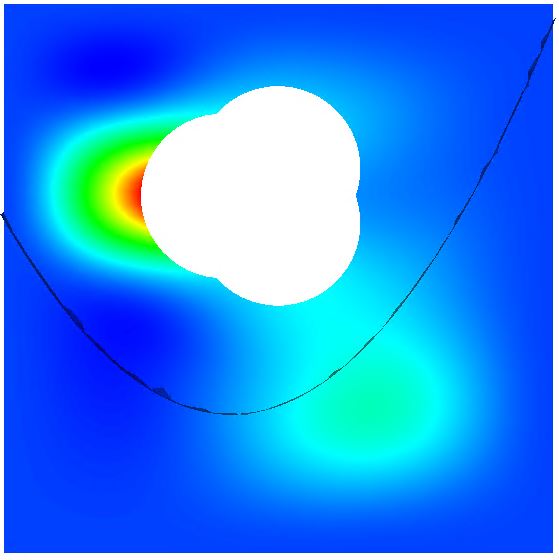} & 1.085\\
\endfirsthead
\endhead
\end{longtable}
\caption{First five eigenmode shape plots for a plate with curvilinear stiffener using a LSM and ABAQUS, a commercial
available software.}
\label{fig:ls_modeplotcomp1}
\end{figure}

To check the accuracy of the level set algorithm, results for clamped curvilinearly stiffened composite plate with a clover shaped cutout are compared with those obtained
using ABAQUS. The curvilinear stiffener configuration was adopted from \cite{Tamijani2010} and scaled proportionally to match the dimensions of the plate. From the ABAQUS library, the plate was meshed using 3040 S8R elements and curvilinear stiffener was meshed using 192 elements S8R elements (48 divisions along the arc length and 4 divisions along the depth direction). The IGA plate model was meshed using 4096 elements and the curvilinear stiffener using 32 elements. The first five eigen buckling models were extracted for each case. The buckling mode shapes obtained using the IGA approach
are similar to those obtained using ABAQUS as can be seen from Figure \ref{fig:ls_modeplotcomp1}.

\FloatBarrier  \FloatBarrier

\FloatBarrier  \FloatBarrier

\section{Advantages and limitations}
\begin{figure}[htbp]
 \centering
  \subfigure[circular]{\includegraphics[width=2.5in]{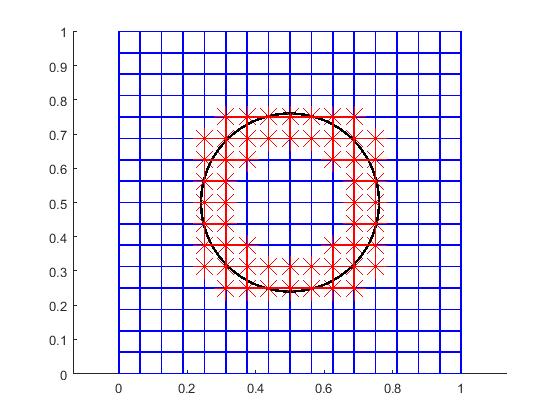}}
  \subfigure[angled elliptical]{\includegraphics[width=2.5in]{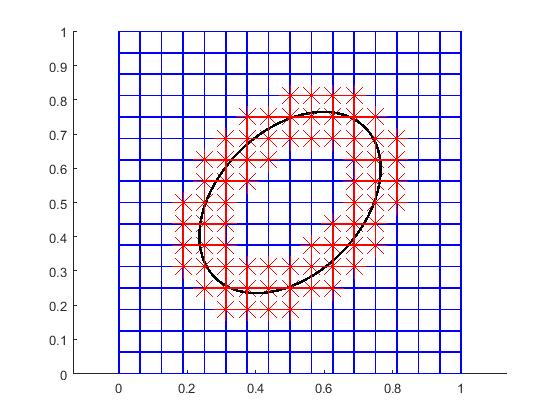}}

    		\caption{Identical IGA mesh for circular and elliptical cutouts}\label{fig:ls_meshplot}
\end{figure}

One of the advantages of the level set method is their ability to use the same mesh controls on the plate for different types of cutouts. Figure \ref{fig:ls_meshplot} shows how the plate is discretized using  $16$ x $16$ elements for circular as well as elliptical cutouts.

\FloatBarrier  \FloatBarrier

%\FloatBarrier \begin{figure}[htbp]	

%	\begin{subfigure}[t]{3in}
%		\centering
%		\includegraphics[width=3in]{circularmesh.jpg}
%		\caption{circular}		
%	\end{subfigure}
%	\quad
%	\begin{subfigure}[t]{3in}
%		\centering
%		\includegraphics[width=3in]{ellipsemesh45.jpg}
%		\caption{angled elliptical}
%	\end{subfigure}
 %   		\caption{Identical IGA mesh for circular and %elliptical cutouts}\label{fig:ls_meshplot}
%\end{figure} \FloatBarrier

Discussions in Section \ref{ss:Appendix} describe the creation of additional degrees of freedom while implementing the level set method. Table \ref{tab:ls_adddof} shows the number of additional degrees of freedom created for each levels of refinement for a plate with a cutout of radius $0.15$. It can be observed that the total degrees of freedom (\textbf{nDoF} = Total number of Gauss points x 5 ) increase much steeply than the additional degrees of freedom (\textbf{Extra DoF}) with increase in refinement levels.

\FloatBarrier \begin{table}[htbp]
\centering
\caption {Additional degrees of freedom with respect to an increase in refinement} \label{tab:ls_adddof} 
\begin{tabular}{|c|c|c|}
\hline
\textbf{Refinement} & \textbf{nDoF} & \textbf{Extra DoF} \\ \hline
3 & 660 & 160 \\ \hline
4 & 1920 & 300 \\ \hline
5 & 6320 & 540 \\ \hline
\end{tabular}
\end{table} \FloatBarrier

Next, computational cost due to these additional degrees of freedom was analyzed. Two cases were considered, a stiffened plate without any cutout (Case 1) and a stiffened plate with a central circular cutout of radius $0.15$. The material properties, ply orientation and stiffener profile were the same as mentioned in Section \ref{ss:ls_noncencutout}. The computational time (in seconds) to perform the static  and the eigenvalue analyses are compared for each case (Table \ref{tab:ls_comptimetab}, Figures \ref{fig:ls_comptimeplotstatic} and \ref{fig:ls_comptimeploteigen}). It is observed that the additional degrees of freedom have a much higher influence on the results from a static analysis as compared to the buckling analysis. However, the difference is very little for lower levels of refinement. 
\FloatBarrier \begin{table}[htbp]
\centering
\caption{Computational time (in seconds) with respect to refinement for static and eigenvalue analyses}
\begin{tabular}{|c|c|c|c|c|c|c|}
\hline
Refinement          & \multicolumn{2}{c|}{4} & \multicolumn{2}{c|}{5} & \multicolumn{2}{c|}{6} \\ \hline
\textbf{}           & Static   & Eigenvalue  & Static   & Eigenvalue  & Static   & Eigenvalue  \\ \hline

Case 1   & 0.0342   & 0.2045      & 0.2008   & 0.7087      & 1.2722   &  4.3813    \\ \hline
Case 2      & 0.0504   & 0.2244      & 0.2014   & 0.8724      & 4.5132    &  4.5352      \\ \hline
\end{tabular}
\label{tab:ls_comptimetab}
\end{table} \FloatBarrier

\FloatBarrier \begin{figure}[htbp]
\centering
\begin{tikzpicture}[scale=1.4]
\tikzstyle{every node}=[font=\small]
\begin{axis}[xmin=3, xmax=6,
ymin=0, ymax=5,
xlabel={Refinement},
ylabel={Time in seconds},legend style={nodes={scale=0.5, transform shape}}]
\addplot coordinates {
(	3	,	0.1075	)
(	4	,	0.2045	)
(	5	,	0.7087	)
(	6	,	4.5352	)

};

\addplot coordinates {
(	3	,	0.1140	)
(	4	,	0.2244	)
(	5	,	0.8724	)
(	6	,	4.3813	)

};
\legend{Simple plate, Plate with a cutout}
\end{axis}
\end{tikzpicture}
\caption{Variation in computational time (in seconds) with respect to refinement for eigenvalue analysis} \label{fig:ls_comptimeploteigen}
\end{figure}
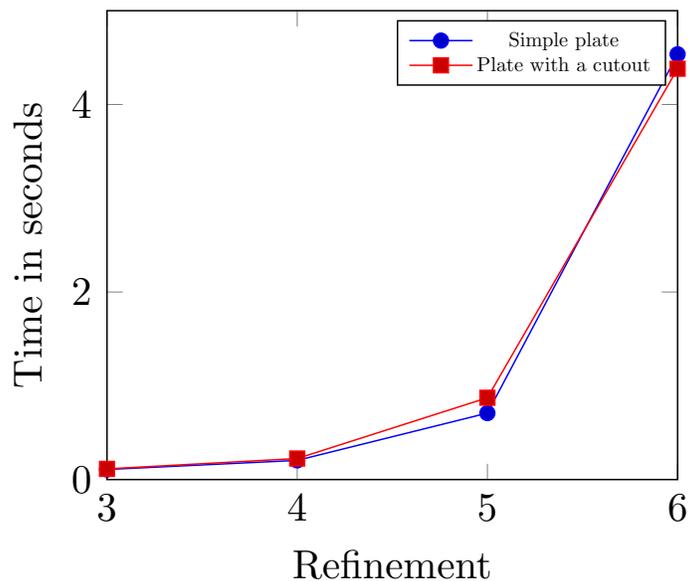 \FloatBarrier

\FloatBarrier \begin{figure}[htbp]
\centering
\begin{tikzpicture}[scale=1.4]
\tikzstyle{every node}=[font=\small]
\begin{axis}[xmin=3, xmax=6,
ymin=0, ymax=5,
xlabel={Refinement},
ylabel={Time in seconds},legend style={nodes={scale=0.5, transform shape}}]
\addplot coordinates {
(	3	,	0.0122	)
(	4	,	0.0342	)
(	5	,	0.2008	)
(	6	,	1.2722	)
};

\addplot coordinates {
(	3	,	0.0127	)
(	4	,	0.0504	)
(	5	,	0.2014	)
(	6	,	4.5132	)
};
\legend{Simple plate, Plate with a cutout}
\end{axis}
\end{tikzpicture}
\caption{Variation in computational time (in seconds) with respect to refinement for static analysis} \label{fig:ls_comptimeplotstatic}
\end{figure}
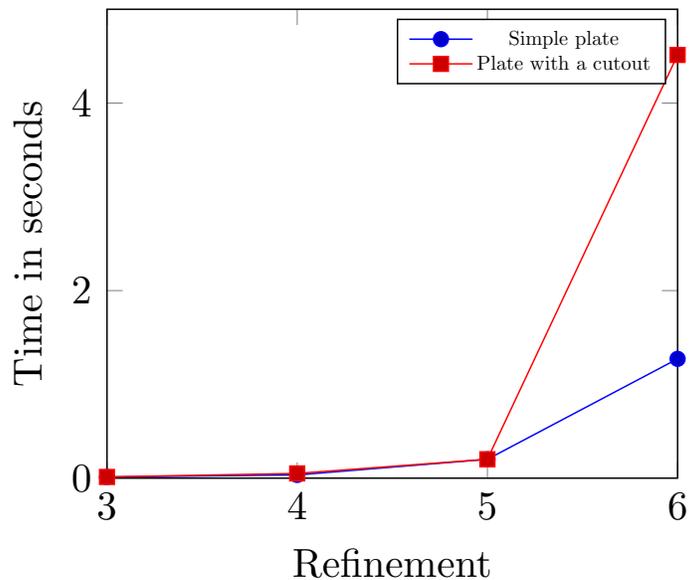 \FloatBarrier

\section{Convergence studies}
Convergence studies on a stiffened plate with a cutout are not as straightforward as a simple plate problem since the size of the cutout, size of the stiffener and the plate meshes all affect the results. We consider a stiffened plate with a central circular cutout. The material properties, ply orientation and stiffener profile were the same as mentioned in Section \ref{ss:ls_noncencutout}. For convergence studies with respect to the increasing cutout size, an unstiffened plate was considered. It can be observed from Figure \ref{fig:ls_varrad} that the rate of convergence becomes slower with an increase in the cutout radius.

\FloatBarrier 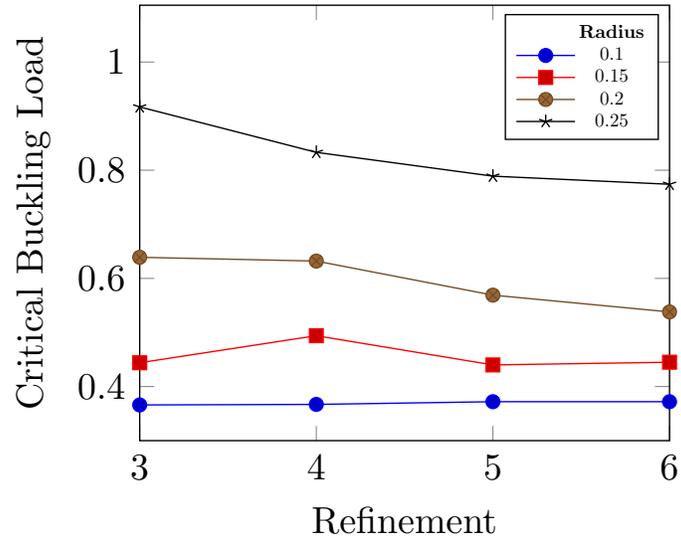
\begin{figure}[htbp]
\centering
\begin{tikzpicture}[scale=1.3]
\tikzstyle{every node}=[font=\small]
\begin{axis}[xmin=3, xmax=6,
ymin=0.3, ymax=1.105,
xlabel={Refinement},
ylabel={Critical Buckling Load},legend style={nodes={scale=0.5, transform shape}}]
\addlegendimage{empty legend}
\addplot coordinates {
(	3	,	.366 	)
(	4	,	.367 	)
(	5	,	.372 	)
(	6	,	.372 	)

};
\addplot coordinates{
(	3,	    .444 	)
(	4	,	.494 	)
(	5	,	.440 	)
(	6	,	.445 	)

};
\addplot coordinates{
(	3,	    .639 	)
(	4	,	.632 	)
(	5	,	.569 	)
(	6	,	.538 	)

};

\addplot coordinates{
(	3	,	.917	)
(	4	,	.833	)
(	5	,	.789	)
(	6	,	.774	)

};

   \addlegendentry{\textbf{Radius}}
   \addlegendentry{0.1}
   \addlegendentry{0.15}
   \addlegendentry{0.2}
   \addlegendentry{0.25}
\end{axis}
\end{tikzpicture}
\caption{Variation of critical buckling load with respect to various plate mesh sizes for cutouts of increasing radii} \label{fig:ls_varrad}
\end{figure} \FloatBarrier

Figures \ref{fig:ls_pointtwofive}, \ref{fig:ls_pointtwo}  and \ref{fig:ls_pointonefive} present the change of critical buckling loads with respect to various stiffener and plate mesh sizes for cutouts of varying radii. It is observed that for cutouts with relatively smaller radius ($0.15$ and $0.2$), for each refinement levels of the plate mesh, the results  converge  for the $4^{th}$ order of refinement of the stiffener mesh. It is also observed that the plate mesh density has a higher influence on the results compared to the stiffener mesh density. However, as the radius of the cutout is increased to $0.25$, the stiffener mesh density starts to impact the convergence of results to a much higher degree (Figure \ref{fig:ls_pointtwofive}).

\FloatBarrier \begin{figure}[htbp]
\centering
\begin{tikzpicture}[scale=1.3]
\tikzstyle{every node}=[font=\small]
\begin{axis}[xmin=2, xmax=8,
ymin=0.87, ymax=.95,
xlabel={Refinement (Stiffener)},
ylabel={Critical Buckling Load},legend style={nodes={scale=0.5, transform shape}}]
\addlegendimage{empty legend}
\addplot coordinates {
(	2	,	0.931198735	)
(	3	,	0.891956611	)
(	4	,	0.90952692	)
(	5	,	0.917617397	)
(	6	,	0.921131239	)
(	7	,	0.922401887	)
(	8	,	0.922837129	)

};
\addplot coordinates{
(	2	,	0.879876197	)
(	3	,	0.890907112	)
(	4	,	0.905919255	)
(	5	,	0.911640979	)
(	6	,	0.909071889	)
(	7	,	0.913829646	)
(	8	,	0.914678852	)

};
\addplot coordinates{

(	2	,	0.896492585	)
(	3	,	0.89218914	)
(	4	,	0.901352978	)
(	5	,	0.905315116	)
(	6	,	0.908086177	)
(	7	,	0.909093211	)
(	8	,	0.909446898	)
};

   \addlegendentry{\textbf{Refinement (Plate)}}
   \addlegendentry{4}
   \addlegendentry{5}
   \addlegendentry{6}
\end{axis}
\end{tikzpicture}
\caption{Variation of critical buckling load with respect to various stiffener and plate mesh sizes for $radius=0.25$} \label{fig:ls_pointtwofive}
\end{figure}
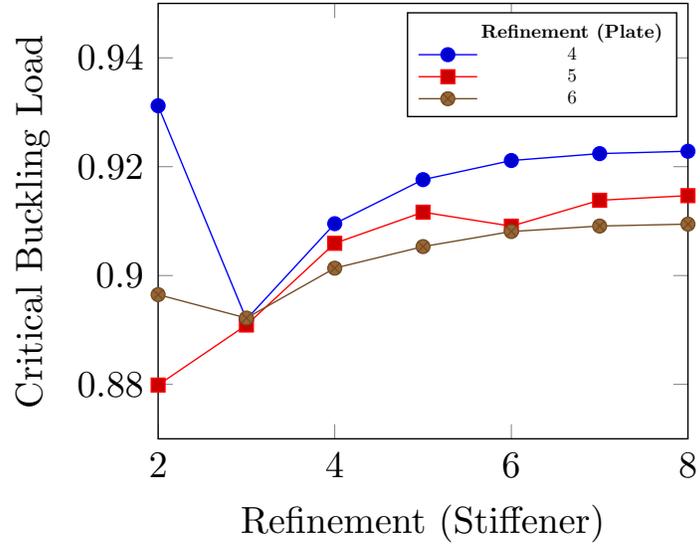 \FloatBarrier

\FloatBarrier 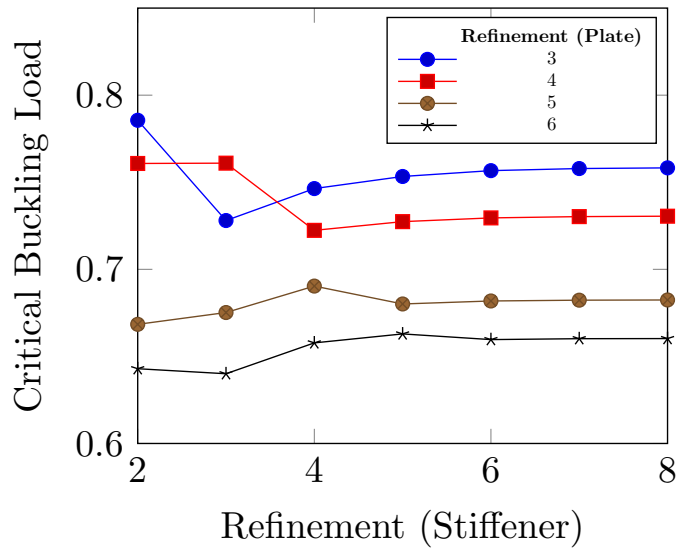
\begin{figure}[htbp]
\centering
\begin{tikzpicture}[scale=1.3]
\tikzstyle{every node}=[font=\small]
\begin{axis}[xmin=2, xmax=8,
ymin=0.6, ymax=.85,
xlabel={Refinement (Stiffener)},
ylabel={Critical Buckling Load},legend style={nodes={scale=0.5, transform shape}}]
\addlegendimage{empty legend}
\addplot coordinates {
(	2	,	0.785585866	)
(	3	,	0.728014764	)
(	4	,	0.746365581	)
(	5	,	0.753312704	)
(	6	,	0.756663008	)
(	7	,	0.757868014	)
(	8	,	0.75825068	)

};
\addplot coordinates{
(	2	,	0.760780547	)
(	3	,	0.760957122	)
(	4	,	0.722379775	)
(	5	,	0.727382264	)
(	6	,	0.729523614	)
(	7	,	0.730265578	)
(	8	,	0.730502365	)

};
\addplot coordinates{
(	2	,	0.668411443	)
(	3	,	0.675191709	)
(	4	,	0.690374475	)
(	5	,	0.680100765	)
(	6	,	0.681821998	)
(	7	,	0.682279132	)
(	8	,	0.682397554	)

};

\addplot coordinates{
(	2	,	0.642922816	)
(	3	,	0.640038601	)
(	4	,	0.657758067	)
(	5	,	0.662905529	)
(	6	,	0.659649271	)
(	7	,	0.660166887	)
(	8	,	0.66023266	)

};

   \addlegendentry{\textbf{Refinement (Plate)}}
   \addlegendentry{3}
   \addlegendentry{4}
   \addlegendentry{5}
   \addlegendentry{6}
\end{axis}
\end{tikzpicture}
\caption{Variation of critical buckling load with respect to various stiffener and plate mesh sizes for $radius=0.2$} \label{fig:ls_pointtwo}
\end{figure} \FloatBarrier

\FloatBarrier 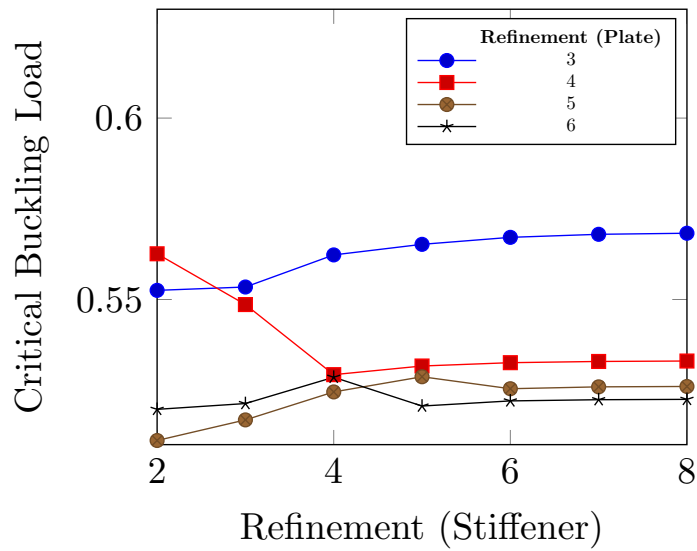
\begin{figure}[htbp]
\centering
\begin{tikzpicture}[scale=1.3]
\tikzstyle{every node}=[font=\small]
\begin{axis}[xmin=2, xmax=8,
ymin=0.51, ymax=.63,
xlabel={Refinement (Stiffener)},
ylabel={Critical Buckling Load},legend style={nodes={scale=0.5, transform shape}}]
\addlegendimage{empty legend}
\addplot coordinates {
(	2	,	0.552518978	)
(	3	,	0.553441739	)
(	4	,	0.562287185	)
(	5	,	0.565204292	)
(	6	,	0.567130821	)
(	7	,	0.567968164	)
(	8	,	0.568261204	)

};
\addplot coordinates{
(	2	,	0.562598107	)
(	3	,	0.548607017	)
(	4	,	0.529286179	)
(	5	,	0.531669178	)
(	6	,	0.532584931	)
(	7	,	0.532922981	)
(	8	,	0.533042715	)

};
\addplot coordinates{
(	2	,	0.511140663	)
(	3	,	0.516815419	)
(	4	,	0.52452539	)
(	5	,	0.528707073	)
(	6	,	0.525408327	)
(	7	,	0.525924593	)
(	8	,	0.526029719	)

};

\addplot coordinates{
(	2	,	0.51972415	)
(	3	,	0.521323703	)
(	4	,	0.528572541	)
(	5	,	0.520636876	)
(	6	,	0.522050067	)
(	7	,	0.522381055	)
(	8	,	0.522474767	)

};

   \addlegendentry{\textbf{Refinement (Plate)}}
   \addlegendentry{3}
   \addlegendentry{4}
   \addlegendentry{5}
   \addlegendentry{6}
\end{axis}
\end{tikzpicture}
\caption{Variation of critical buckling load with respect to various stiffener and plate mesh sizes for $radius=0.15$} \label{fig:ls_pointonefive}
\end{figure} \FloatBarrier
\section{Conclusions}
This paper discusses the level set method which requires very little computational effort to obtain a mesh independent description of the geometrical shape of a complicated cutout.
The level sets discussed in this paper are implemented in the
form of a signed distance function which is both computationally inexpensive and easy to
implement. Such an implementation is greatly useful since the tensor product nature of the NURBS basis functions only allows for global refinement and therefore treating the trimmed objects such as internal cutouts is rather difficult. However, implementing the level set method obviates the use of trimmed NURBS surface to describe complex geometrical features like the cutouts. The efficiency of this method for thermal buckling analysis is described with the help of numerical examples for stiffened laminated plates containing complicated cutouts of various shapes and dimensions.
\newline
\section{Acknowledgement}
The authors would like to thank the Institute for Critical Technology and Applied Science (ICTAS) at Virginia Tech for providing financial support which made this research possible.
\newpage
\section*{References}
\bibliography{mybibfile}

\begin{thebibliography}{10}
\expandafter\ifx\csname url\endcsname\relax
  \def\url#1{\texttt{#1}}\fi
\expandafter\ifx\csname urlprefix\endcsname\relax\def\urlprefix{URL }\fi
\expandafter\ifx\csname href\endcsname\relax
  \def\href#1#2{#2} \def\path#1{#1}\fi

\bibitem{Hughes2005}
T.~J. Hughes, J.~A. Cottrell, Y.~Bazilevs, {Isogeometric analysis: CAD, finite
  elements, NURBS, exact geometry and mesh refinement}, Computer Methods in
  Applied Mechanics and Engineering 194~(39-41) (2005) 4135--4195.

\bibitem{Yu2015}
T.~T. Yu, S.~Yin, T.~Q. Bui, S.~Hirose, {A simple FSDT-based isogeometric
  analysis for geometrically nonlinear analysis of functionally graded plates},
  Finite Elements in Analysis and Design 96~(C) (2015) 1--10.

\bibitem{de2018structural}
S.~De, K.~Singh, B.~Alanbay, R.~K. Kapania, R.~Aguero, Structural optimization
  of truck front-frame under multiple load cases, in: ASME International
  Mechanical Engineering Congress and Exposition, Vol. 52187, American Society
  of Mechanical Engineers, 2018, p. V013T05A039.

\bibitem{de2019structural}
S.~De, K.~Singh, J.~Seo, R.~K. Kapania, E.~Ostergaard, N.~Angelini, R.~Aguero,
  Structural design and optimization of commercial vehicles chassis under
  multiple load cases and constraints, in: AIAA Scitech 2019 Forum, 2019, p.
  0705.

\bibitem{de2021lightweight}
S.~De, K.~Singh, J.~Seo, R.~K. Kapania, E.~Ostergaard, N.~Angelini, R.~Aguero,
  Lightweight chassis design of hybrid trucks considering multiple road
  conditions and constraints, World Electric Vehicle Journal 12~(1) (2021) 3.

\bibitem{de2019unconventional}
S.~De, K.~Singh, J.~Seo, R.~Kapania, R.~Aguero, E.~Ostergaard, N.~Angelini,
  Unconventional truck chassis design with multi-functional cross members,
  Tech. rep., SAE Technical Paper (2019).

\bibitem{jrad2017global}
M.~Jrad, S.~De, R.~K. Kapania, Global-local aeroelastic optimization of
  internal structure of transport aircraft wing, in: 18th AIAA/ISSMO
  Multidisciplinary Analysis and Optimization Conference, 2017, p. 4321.

\bibitem{de2021algorithms}
S.~De, R.~K. Kapania, Algorithms for 2d mesh decomposition in distributed
  design optimization (2021).
\newblock \href {http://arxiv.org/abs/2002.00525} {\path{arXiv:2002.00525}}.

\bibitem{Bazilevs2008}
Y.~Bazilevs, V.~M. Calo, T.~J.~R. Hughes, Y.~Zhang, {Isogeometric
  fluid-structure interaction: theory, algorithms, and computations},
  Computational Mechanics 43~(1) (2008) 3--37.

\bibitem{Evans2009a}
J.~A. Evans, Y.~Bazilevs, I.~Babu{\v{s}}ka, T.~J. Hughes, {$n$-Widths,
  $sup$-infs, and optimality ratios for the $k$-version of the isogeometric
  finite element method}, Computer Methods in Applied Mechanics and Engineering
  198~(21-26) (2009) 1726--1741.

\bibitem{Hughes2010a}
T.~J.~R. Hughes, A.~Reali, G.~Sangalli, {Efficient quadrature for NURBS-based
  isogeometric analysis}, Computer Methods in Applied Mechanics and Engineering
  199~(5-8) (2010) 301--313.

\bibitem{Auricchio2012a}
F.~Auricchio, F.~Calabr{\`{o}}, T.~J. Hughes, A.~Reali, G.~Sangalli, {A simple
  algorithm for obtaining nearly optimal quadrature rules for NURBS-based
  isogeometric analysis}, Computer Methods in Applied Mechanics and Engineering
  249-252 (2012) 15--27.

\bibitem{Bazilevs2010}
Y.~Bazilevs, C.~Michler, V.~M. Calo, T.~J. Hughes, {Isogeometric variational
  multiscale modeling of wall-bounded turbulent flows with weakly enforced
  boundary conditions on unstretched meshes}, Computer Methods in Applied
  Mechanics and Engineering 199~(13-16) (2010) 780--790.

\bibitem{Valizadeh2013}
N.~Valizadeh, T.~Bui, V.~Vu, H.~Thai, M.~Nguyen, {Isogeometric Simulation for
  Buckling, Free and Forced Vibration of Orthotropic Plates}, International
  Journal of Applied Mechanics 05~(02) (2013) 1350017.

\bibitem{yin2014isogeometric}
S.~Yin, J.~S. Hale, T.~Yu, T.~Q. Bui, S.~P. Bordas, {Isogeometric locking-free
  plate element: A simple first order shear deformation theory for functionally
  graded plates}, Composite Structures 118~(1) (2014) 121--138.

\bibitem{kapoor2013interlaminar}
H.~Kapoor, R.~K. Kapania, S.~R. Soni, {Interlaminar stress calculation in
  composite and sandwich plates in NURBS Isogeometric finite element analysis},
  Composite Structures 106 (2013) 537--548.

\bibitem{kapoor2012geometrically}
H.~Kapoor, R.~K. Kapania, {Geometrically nonlinear NURBS isogeometric finite
  element analysis of laminated composite plates}, {Composite Structures}
  94~(12) (2012) 3434--3447.

\bibitem{Verhoosel2011a}
C.~V. Verhoosel, M.~A. Scott, T.~J.~R. Hughes, R.~de~Borst, {An isogeometric
  analysis approach to gradient damage models}, International Journal for
  Numerical Methods in Engineering 86~(1) (2011) 115--134.
\newblock \href {http://arxiv.org/abs/1010.1724} {\path{arXiv:1010.1724}}.

\bibitem{Lu2011}
J.~Lu, {Isogeometric contact analysis: Geometric basis and formulation for
  frictionless contact}, Computer Methods in Applied Mechanics and Engineering
  200~(5-8) (2011) 726--741.

\bibitem{nguyen2014isogeometric}
H.~Nguyen-Xuan, L.~V. Tran, C.~H. Thai, S.~Kulasegaram, S.~P. Bordas,
  {Isogeometric analysis of functionally graded plates using a refined plate
  theory}, Composites Part B: Engineering 64 (2014) 222--234.

\bibitem{Wall2008}
W.~A. Wall, M.~A. Frenzel, C.~Cyron, {Isogeometric structural shape
  optimization}, Computer Methods in Applied Mechanics and Engineering
  197~(33-40) (2008) 2976--2988.

\bibitem{kiendl2009isogeometric}
J.~Kiendl, K.-U. Bletzinger, J.~Linhard, R.~W{\"u}chner, {Isogeometric shell
  analysis with Kirchhoff--Love elements}, Computer Methods in Applied
  Mechanics and Engineering 198~(49-52) (2009) 3902--3914.

\bibitem{Kiendl2010}
J.~Kiendl, Y.~Bazilevs, M.~C. Hsu, R.~W{\"{u}}chner, K.~U. Bletzinger, {The
  bending strip method for isogeometric analysis of Kirchhoff-Love shell
  structures comprised of multiple patches}, Computer Methods in Applied
  Mechanics and Engineering 199~(37-40) (2010) 2403--2416.

\bibitem{shojaee2012free}
S.~Shojaee, E.~Izadpanah, N.~Valizadeh, J.~Kiendl, {Free vibration analysis of
  thin plates by using a NURBS-based isogeometric approach}, Finite Elements in
  Analysis and Design 61 (2012) 23--34.

\bibitem{tran2014isogeometric}
L.~V. Tran, C.~H. Thai, H.~T. Le, B.~S. Gan, J.~Lee, H.~Nguyen-Xuan,
  Isogeometric analysis of laminated composite plates based on a four-variable
  refined plate theory, Engineering Analysis with Boundary Elements 47 (2014)
  68--81.

\bibitem{Piegl1998}
L.~A. Piegl, W.~Tiller, {Geometry-based triangulation of trimmed NURBS
  surfaces}, Computer-Aided Design 30~(1) (1998) 11--18.

\bibitem{Shojaee2012a}
S.~Shojaee, N.~Valizadeh, E.~Izadpanah, T.~Bui, T.~V. Vu, {Free vibration and
  buckling analysis of laminated composite plates using the NURBS-based
  isogeometric finite element method}, Composite Structures 94~(5) (2012)
  1677--1693.

\bibitem{Schmidt2012}
R.~Schmidt, R.~W{\"{u}}chner, K.~U. Bletzinger, {Isogeometric analysis of
  trimmed NURBS geometries}, Computer Methods in Applied Mechanics and
  Engineering 241-244 (2012) 93--111.

\bibitem{Ghorashi2012a}
S.~S. Ghorashi, N.~Valizadeh, S.~Mohammadi, {Extended isogeometric analysis for
  simulation of stationary and propagating cracks}, International Journal for
  Numerical Methods in Engineering 89~(9) (2012) 1069--1101.
\newblock \href {http://arxiv.org/abs/1010.1724} {\path{arXiv:1010.1724}}.

\bibitem{Rank2012}
E.~Rank, M.~Ruess, S.~Kollmannsberger, D.~Schillinger, A.~D{\"{u}}ster,
  {Geometric modeling, isogeometric analysis and the finite cell method},
  Computer Methods in Applied Mechanics and Engineering 249-252 (2012)
  104--115.

\bibitem{Schillinger2012a}
D.~Schillinger, L.~Ded{\`{e}}, M.~A. Scott, J.~A. Evans, M.~J. Borden, E.~Rank,
  T.~J. Hughes, {An isogeometric design-through-analysis methodology based on
  adaptive hierarchical refinement of NURBS, immersed boundary methods, and
  T-spline CAD surfaces}, Computer Methods in Applied Mechanics and Engineering
  249-252 (2012) 116--150.

\bibitem{devarajan2016thermomechanical}
B.~Devarajan, D.~Locatelli, R.~Kapania, R.~Meritt, Thermo-mechanical analysis
  and design of threaded fasteners, in: 2016 AIAA/ASCE/AHS/ASC Structures,
  Structural Dynamics, and Materials Conference, 2016, p. 0579.

\bibitem{2016-01-1986}
Q.~Li, B.~Devarajan, X.~Zhang, R.~Burgos, D.~Boroyevich, P.~Raj, Conceptual
  design and weight optimization of aircraft power systems with high-peak
  pulsed power loads, in: SAE Technical Paper, SAE International, 2016, p.~10.
\newblock \href {http://dx.doi.org/10.4271/2016-01-1986}
  {\path{doi:10.4271/2016-01-1986}}.

\bibitem{noor1996computational}
A.~K. Noor, W.~S. Burton, C.~W. Bert, {Computational Models for Sandwich Panels
  and Shells}, Applied Mechanics Reviews 49~(3) (1996) 155.

\bibitem{Burton1994}
W.~S. Burton, A.~K. Noor, {Three Dimensional Solutions for Thermomechanical
  Stresses in Sandwich Panels and Shells}, Journal of Engineering Mechanics
  120~(10) (1994) 2044--2071.

\bibitem{noor1992three}
A.~K. Noor, W.~S. Burton, {Three Dimensional Solutions for Thermal Buckling of
  Multilayered Anisotropic Plates}, Journal of Engineering Mechanics 118~(4)
  (1992) 683--701.

\bibitem{mukhopadhyay1990finite}
M.~Mukhopadhyay, A.~Mukherjee, {Finite element buckling analysis of stiffened
  plates}, Computers \& structures 34~(6) (1990) 795--803.

\bibitem{prusty2001finite}
B.~G. Prusty, S.~Satsangi, {Finite element transient dynamic analysis of
  laminated stiffened shells}, Journal of Sound and Vibration 248~(2) (2001)
  215--233.

\bibitem{Tamijani2010}
A.~Y. Tamijani, R.~K. Kapania, {Buckling and Static Analysis of Curvilinearly
  Stiffened Plates Using Mesh-Free Method}, {AIAA Journal} 48~(12) (2010)
  2739--2751.

\bibitem{Zhao2016}
W.~Zhao, R.~K. Kapania, {Buckling analysis of unitized curvilinearly stiffened
  composite panels}, Composite Structures 135 (2016) 365--382.

\bibitem{robinson2016aeroelastic}
J.~H. Robinson, S.~Doyle, G.~Ogawa, M.~Baker, S.~De, M.~Jrad, R.~K. Kapania,
  Aeroelastic optimization of wing structure using curvilinear spars and ribs
  (sparibs), in: 17th AIAA/ISSMO Multidisciplinary Analysis and Optimization
  Conference, 2016, p. 3994.

\bibitem{de2017sparibs}
S.~De, M.~Jrad, D.~Locatelli, R.~K. Kapania, M.~Baker, Sparibs geometry
  parameterization for wings with multiple sections using single design space,
  in: 58th AIAA/ASCE/AHS/ASC Structures, Structural Dynamics, and Materials
  Conference, 2017, p. 0570.

\bibitem{de2017structural}
S.~De, Structural modeling and optimization of aircraft wings having
  curvilinear spars and ribs (sparibs), Ph.D. thesis, Virginia Tech (2017).

\bibitem{Shi2015}
P.~Shi, R.~K. Kapania, C.~Y. Dong, {Vibration and Buckling Analysis of
  Curvilinearly Stiffened Plates Using Finite Element Method}, {AIAA Journal}
  53~(5) (2015) 1319--1335.

\bibitem{hao2016optimization}
P.~Hao, B.~Wang, K.~Tian, G.~Li, X.~Zhang, {Optimization of curvilinearly
  stiffened panels with single cutout concerning the collapse load},
  International Journal of Structural Stability and Dynamics 16~(07) (2016)
  1550036.

\bibitem{devarajan2020thermal}
B.~Devarajan, R.~K. Kapania, Thermal buckling of curvilinearly stiffened
  laminated composite plates with cutouts using isogeometric analysis,
  Composite Structures (2020) 111881.

\bibitem{devarajan2019thermomechanical}
B.~Devarajan, Thermomechanical and Vibration Analysis of Stiffened Unitized
  Structures and Threaded Fasteners, vtechworks.lib.vt.edu, 2019.

\bibitem{Piegl1996}
L.~L. Piegl, W.~Tiller, {The NURBS Book}, Vol.~28, Springer Science {\&}
  Business Media, 1996.
\newblock \href {http://arxiv.org/abs/arXiv:1011.1669v3}
  {\path{arXiv:arXiv:1011.1669v3}}.

\bibitem{Tran2017}
L.~V. Tran, M.~A. Wahab, S.~E. Kim, {An isogeometric finite element approach
  for thermal bending and buckling analyses of laminated composite plates},
  Composite Structures 179 (2017) 35--49.

\bibitem{babu2000refined}
C.~{Sarath Babu}, T.~Kant, {Refined higher order finite element models for
  thermal buckling of laminated composite and sandwich plates}, Journal of
  Thermal Stresses 23~(2) (2000) 111--130.

\bibitem{Kant2000}
T.~Kant, C.~S. Babu, {Thermal buckling analysis of skew fibre-reinforced
  composite and sandwich plates using shear deformable finite element models},
  Composite Structures 49~(1) (2000) 77--85.

\bibitem{Avci2005}
A.~Avci, S.~Kaya, B.~Daghan, {Thermal buckling of rectangular laminated plates
  with a hole}, Journal of Reinforced Plastics and Composites 24~(3) (2005)
  259--272.

\end{thebibliography}

\end{document}